\documentclass[trackchanges]{aastex701}

\usepackage{natbib}
\usepackage{comment}
\usepackage{subcaption}

\begin{document}

\title{Exploring the long-term temporal variability in polarization through multi-epoch optical spectro-polarimetry - Part II: A sample of symbiotic and red-giant stars}

\author[orcid=0000-0001-7340-8873,sname='Maiti']{Arijit Maiti}
\affiliation{Astronomy \& Astrophysics Division, Physical Research Laboratory, Ahmedabad 380009, Gujarat, India}
\affiliation{Indian Institute of Technology Gandhinagar, Gandhinagar, India, 382055}
\email[show]{arijitmaiti@prl.res.in}  

\author[orcid=0000-0002-6222-3045,sname='Pandey']{Ruchi Pandey} 
\altaffiliation{Current affiliation: Department of Physics \& Astronomy, Johns Hopkins University, 3400 N. Charles St, Baltimore, MD 21218, United States}
\altaffiliation{Current affiliation: X-ray Astrophysics Laboratory, NASA Goddard Space Flight Center, Code 662, Greenbelt, MD 20771, United States}
\affiliation{Astronomy \& Astrophysics Division, Physical Research Laboratory, Ahmedabad 380009, Gujarat, India}
\email[show]{ruchi.pandey@nasa.gov}

\author[gname=Sube Singh,sname='Gurjar']{Sube Singh Gurjar} 
\affiliation{Astronomy \& Astrophysics Division, Physical Research Laboratory, Ahmedabad 380009, Gujarat, India}
\affiliation{Indian Institute of Technology Gandhinagar, Gandhinagar, India, 382055}
\email[]{}

\author[gname=Mudit K.,sname='Srivastava']{Mudit K. Srivastava} 
\affiliation{Astronomy \& Astrophysics Division, Physical Research Laboratory, Ahmedabad 380009, Gujarat, India}
\email[show]{mudit@prl.res.in}

\begin{abstract}

Measuring polarization and its temporal variability in the continuum and across emission features provides a powerful probe of the small-scale circumstellar environments of astrophysical sources, on spatial scales otherwise inaccessible to direct imaging techniques. However, the photon-hungry nature of spectro-polarimetry has significantly limited the availability of such datasets in the literature. Here we present the results of a multi-epoch spectro-polarimetric survey of a sample of symbiotic and red giant stars. These observations were first conducted during the performance verification phase of a recently developed medium-resolution echelle spectro-polarimeter, named ProtoPol, mounted on the Physical Research Laboratory (PRL) 2.5m telescope, Mt Abu, India. Considering the scarcity of such observing samples in the literature, the targets were re-observed over a period exceeding 26 months, from March 2024 to May 2026, in order to explore the time variability of their polarimetric measurements. Our sample comprises 6 symbiotic stars and 18 red giant stars. Our observations found that, unlike the Raman-scattered $\lambda \lambda$ $6830, 7088 \text{\AA}$ features, the H$\alpha$ emission in most of the symbiotics did not exhibit any notable polarization signature. However, the continuum polarization of most symbiotics and red giants in the sample showed significant variation between observation epochs. The observations presented here constitute one of the rare multi-epoch spectro-polarimetric datasets spanning more than two years and should be of considerable interest to the broader astronomical community. This paper is Part-II of a two-part series of sample studies; corresponding results for Herbig Ae/Be and classical Be stars are presented in Part-I.

\end{abstract}



\section{Introduction} 
\label{sec:Introduction} 

This is Paper-II of the two-part paper series, reporting the multi-epoch spectro-polarimetric observations of a sample of astrophysical objects observed with ProtoPol, a medium-resolution echelle spectro-polarimeter developed for PRL 2.5m telescope at Mt. Abu, India \citep{kumar2022designs, srivastava2024development, srivastava2026development, maiti2026development}. Developed fully in-house with completely off-the-shelf optical components, ProtoPol operates with a spectral resolution between 0.4-0.75 $\AA$, covering the entire visible wavelength range from 4000-9600 $\AA$. During the on-sky characterization phase of the instrument, a large sample of stars was observed with ProtoPol for the purpose of science verification. The sample consisted of hot stars such as Herbig Ae/Be and classical Be stars, and cool stars such as symbiotic and red giant stars, observed over multiple epochs over a period of around 26 months, since the instrument's first light. The background and rationale behind the current study are presented in the Paper-I of the series, along with the multi-epoch spectro-polarimetric results for the sample of Herbig Ae/Be $\&$ classical Be stars. Here we present the results obtained for a sample of cool stars. 
\par
Spectro-polarimetric observations measure the polarization state of light as a function of wavelength by combining spectroscopy and polarimetry \citep{2003bookDelToro}. While spectroscopy usually provides information on line profiles, velocity structure, and excitation conditions of the astrophysical source, polarimetry is sensitive to the geometry of the emitting and scattering regions. This makes spectro-polarimetry particularly useful for studying circumstellar environments that are too compact to be spatially resolved by direct imaging. The polarization vectors generated by scattering almost completely cancel out if the source geometry is spherically symmetric. However, if the geometry of the source deviates from spherical symmetry, scattering by dust grains, molecules, atoms, or free electrons can result in detectable polarization. Thus, spectro-polarimetric observations provide a powerful indirect method for probing unresolved circumstellar environments, including disks, bipolar outflows, clumpy winds, asymmetric envelopes, and binary interaction regions \citep[see e.g.,][and references therein]{brown1978polarisation, 1977A&ABrown, 2003bookDelToro, 2025GalaxIgnace, 2012AIPCVink}. Evolved stellar systems are ideal targets for spectro-polarimetric studies because they are commonly surrounded by circumstellar material produced by stellar winds, dust formation, pulsation-driven mass loss, and binary interaction \citep{willson2000mass, 2018A&AHofner, munari2019symbiotic}. The composition, optical depth, and distribution of the scattering material, as well as the relative geometry of the illuminating source and the scattering region, all affect the measured polarization in such systems \citep{1977A&ABrown, brown1978polarisation, 2009A&AIgnace}. Thus, wavelength-dependent and time-dependent polarization can provide information on the morphology and temporal evolution of circumstellar material that is difficult to obtain from spectroscopy or photometry alone.
\par
Symbiotic stars (SySt) represent one important class of such evolved systems. They are long-period interacting binary systems composed of an evolved cool giant, either a red giant in S-type systems or a Mira variable surrounded by a dusty envelope in D-type systems, and a hot companion, most commonly a white dwarf (WD) \citep{kenyon1986symbiotic, munari2019symbiotic}. Their interaction gives rise to a characteristic composite spectrum marked by prominent emission lines superimposed on a cool stellar continuum rich in broad molecular absorption features \citep{kenyon1986symbiotic, van1993atlas, ivison1994atlas}. These systems typically exhibit orbital periods ranging from several years to decades. The circumstellar environment of SySt, shaped by mass loss from the giant and by the ionizing radiation and stellar winds of the hot companion, contains both ionized and neutral regions \citep{kenyon1986symbiotic, munari2019symbiotic, schmid1994raman}. Owing to their intrinsically complex environments, SySt serve as important probes of the late evolutionary stages of low- and intermediate-mass stars and provide valuable laboratories for investigating binary interaction processes and stellar evolution. 
\par 
One of the most distinct spectral features of many SySt is the presence of emission features at $\lambda \lambda$ $6830, 7088$ {\AA}, first documented by \cite{allen1980unidentified} in a sample study of SySt, but later realized to be produced by Raman scattering of O VI $\lambda \lambda$ $1032, 1038 $ {\AA} resonance doublet \citep{schmid1989identification,nussbaumer1989raman}. While the resonance doublet originates in the vicinity of the hot component, Raman scattering of the UV photons in the neutral hydrogen wind of the giant gives rise to the $\lambda \lambda$ $6830, 7088 \text{\AA}$ features in the optical spectrum. Nearly 50 \% of the SySt show these features \citep{van1993atlas, ivison1994atlas} and are used as a diagnostic to classify a star as a symbiotic. Apart from the $\lambda \lambda$ $6830, 7088 \text{\AA}$ features, several other Raman scattered features are also observed in the optical spectra of symbiotics, although detection of such features is rarer. One such feature is the Raman scattering of Lyman $\beta$ (Ly$\beta$) photons in the neutral hydrogen wind of the red giant, with the scattered photons falling on the wings of the H$\alpha$ emission line, and giving rise to the typical broad H$\alpha$ wings observed in symbiotics \citep{lee2000raman, yoo2002polarization, ikeda2004polarized}. A recent example is provided by \cite{maiti2026discovery}, who employed multi-epoch spectro-polarimetric observations with ProtoPol to detect orbital phase-dependent variations in the H$\alpha$ polarization signature of the symbiotic star Y Gem. On the other hand, Thomson scattering of H$\alpha$ photons by electrons has also been predicted to cause polarization of the H$\alpha$ emission \citep{kim2007Thompson, chang2018broad}. Polarized H$\alpha$ line in the symbiotic star BI Crucis, detected through spectro-polarimetric observations, was suspected to be caused by Thomson scattering \citep{harries1996accretion}. Thus, SySt provides a rich environment for observing various scattering processes.
\par
Red giant stars form the second class of objects considered in this work. These are evolved low to intermediate-mass stars with extended atmospheres, low surface gravities, pulsation-driven variability, and considerable mass loss. In many situations, particularly for AGB and Mira-type stars, they lose mass at a large rate ($\dot{M} > 10^{-7} M_\odot /\text{yr}$) and at a low velocity (10 km/s), leading to the formation of extended gaseous and dusty circumstellar envelopes \citep{1996A&ARvHabing, willson2000mass, 2018A&AHofner}. Spectro-polarimetric observations of red giant stars provide an important diagnostic of envelope asymmetries. If the circumstellar material is dispersed spherically, the polarisation vectors produced by scattering cancel out, leaving little or no net polarisation. Thus, the identification of intrinsic continuous polarisation shows deviations from spherical symmetry, such as clumpy dust generation, localised mass-loss events, flattened envelopes, or bipolar structures \citep{boyle1986ccd, 2005MNRASIreland, 2012NaturNorris, 2016A&AOhnaka}. Multi-epoch continuum spectro-polarimetry of red giants is particularly useful because pulsation, shocks, convection, dust condensation, and episodic mass loss can modify the scattering geometry on timescales of months to years \citep{willson2000mass, 2018A&AHofner, bieging2006optical}. Temporal changes in the degree of polarization and polarization position angle can therefore reveal the evolving structure of the circumstellar environment \citep{brandi2000linear, bieging2006optical}.
\par
In this work, we present a multi-epoch optical spectro-polarimetric study of 24 evolved stellar objects observed over a period of nearly 26 months, from March 2024 to May 2026. The sample consists of 6 symbiotic stars and 18 red giant stars. Large-sample multi-epoch spectro-polarimetric studies of red giant stars remain limited in the literature, especially for relatively low-amplitude semiregular and irregular variables, for which the circumstellar envelopes may be weak or compact. In this study, we try to address this gap in the literature by presenting multi-epoch optical continuum spectro-polarimetry of these red giant stars, including semiregular, irregular, Mira-type, and AGB candidates. For the symbiotic stars, we investigate both continuum polarization and line polarization across key emission features, with particular emphasis on the H$\alpha$ profile and the Raman-scattered O VI features where present. For the red giant stars, we report the continuum polarization and its temporal variability as a probe of asymmetric and evolving circumstellar envelopes. Several objects in our sample show changes in their polarization properties between epochs, indicating dynamic scattering environments. This dataset therefore provides a valuable multi-epoch spectro-polarimetric view of evolved stellar systems and offers new constraints on the geometry and temporal evolution of their circumstellar material. The paper is organized as follows. Section~\ref{sec:ObservationCampaign} describes the observations, instrumental setup, data reduction, and polarimetric extraction. Section~\ref{sec:results-and-discussion} presents the results and discussion for both the symbiotic-star and red giant samples. Finally, Section~\ref{sec:Summary} summarizes the main findings.


\section{Observations and Data Reduction}
\label{sec:ObservationCampaign}

\subsection{ProtoPol}
Optical spectro-polarimetric observations were conducted using ProtoPol, a medium-resolution echelle spectro-polarimeter \citep{kumar2022designs, srivastava2024development, srivastava2026development, maiti2026development}, currently mounted on PRL 2.5m telescope at the Mt. Abu Observatory, Gurushikhar, India. ProtoPol covers the entire visible wavelength range from 4000-9600 
{\AA} with a spectral resolution of 0.4 - 0.75 {\AA}. The instrument couples a polarimeter unit to a spectrometer unit, which employs echelle and cross-disperser (CD) gratings to produce cross-dispersed spectra of the ordinary and extraordinary beams, hereafter o- and e-rays, for each echelle order on a $1\mathrm{K} \times 1\mathrm{K}$ Andor CCD detector. Two separate CD gratings are used: one for the blue wavelength range, 4000 - 6200 {\AA}, and another for the red wavelength range, 5800 - 9600 {\AA}. The instrument is also equipped with a calibration unit that houses a Uranium-Argon (UAr) lamp for wavelength calibration of the echelle orders and a halogen lamp for order tracing. The on-sky performance and characterization of the ProtoPol instrument are described in detail in \cite{maiti2026development}.

\subsection{Observing Campaign and Sample Selection}

The observing campaign lasted 26 months, from March 2024 to May 2026, following the instrument's first-light observations. The observations were conducted in parallel with the instrument characterization observations. A large sample of symbiotic and red giant stars was observed over multiple epochs, with at least two observations per source, to investigate their multi-epoch spectro-polarimetric variability. In total, 6 symbiotic stars and 18 red giant stars were targeted in this sample study. The observation logs for the symbiotic and red giant samples are provided in Tables~\ref{tab:syst} and ~\ref{tab:redgiant}, respectively. The sample was selected based on the apparent magnitudes of the targets, $V \lesssim 11$ mag for observations with the 2.5 m telescope, the presence of H$\alpha$ emission in the optical spectra for symbiotic stars, and target visibility during the observing campaign. The first-epoch observations of a few stars in the sample were previously reported in \cite{maiti2026development} and are included in the present study for continuity.


\begin{deluxetable*}{l l c c l c c} 
\tablecaption{Observation Log of the Symbiotic Star Sample
\label{tab:syst}} 
\tablehead{ \colhead{Name} & \colhead{Alternate Name} & \colhead{$V$} & \colhead{$G$} & \colhead{Spectral Type} & \colhead{Date} & \colhead{Exposure Time} \\
\colhead{} & \colhead{} & \colhead{(mag)} & \colhead{(mag)} & \colhead{} & \colhead{(dd-mm-yyyy)} & \colhead{(s $\times$ HWP pos. $\times$ sets)} } 
\startdata 
AG Peg & HD 207757 & 8.69 & 7.81 & M3 IIIe C & 27-04-2024 & $600 \times 4 \times 4$ \\ 
& & & & & 19-12-2025 & $900 \times 4 \times 1$ \\ 
RW Hya & HD 117970 & 10.00 & 8.12 & M1/2 IIIe C & 28-04-2024 & $600 \times 4 \times 4$ \\ 
& & & & & 19-01-2026 & $900 \times 4 \times 1$ \\ 
T CrB & HD 143454 & 10.24 & 8.72 & M3 IIIe\_sh C & 25-01-2025 & $1800 \times 4 \times 1$ \\ 
& & & & & 01-03-2025 & $900 \times 4 \times 3$ \\ 
& & & & & 16-01-2026 & $1200 \times 4 \times 1$ \\ 
& & & & & 18-02-2025 & $1800 \times 4 \times 1$ \\ 
UV Aur & HD 34842 & 10.41 & 7.57 & C8,1Je C & 07-03-2024 & $600 \times 4 \times 3$ \\ 
& & & & & 12-04-2026 & $1500 \times 4 \times 1$ \\ 
AG Dra & HIP 78512 & 9.74 & 9.17 & K3 IIIep C & 02-04-2025 & $900 \times 4 \times 2$ \\ 
& & & & & 14-05-2026 & $1800 \times 4 \times 1$ \\ 
Z And & HD 221650 & 8.00 & 9.13 & M2 III+B1eq C & 26-01-2025 & $900 \times 4 \times 1$ \\ 
& & & & & 18-12-2025 & $1800 \times 4 \times 1$ \\ 
\enddata 
\tablenotetext{}{The $V$ and $G$ band magnitudes and spectral types are adopted from SIMBAD. The $G$ magnitudes correspond to Gaia EDR3 values \citep{brown2021gaia}.} 
\end{deluxetable*}



\begin{deluxetable*}{l c c c l l c c}
\tablecaption{Observation Log of the Red Giant Star Sample\label{tab:redgiant}}
\tablehead{
\colhead{Name} &
\colhead{HD Number} &
\colhead{$V$} &
\colhead{$G$} &
\colhead{Spectral Type} &
\colhead{Classification} &
\colhead{Date} &
\colhead{Exposure Time}
\\
\colhead{} &
\colhead{} &
\colhead{(mag)} &
\colhead{(mag)} &
\colhead{} &
\colhead{} &
\colhead{(dd-mm-yyyy)} &
\colhead{(s $\times$ HWP pos.)}
}
\startdata
BK Vir       & 108849 & 7.28 & 4.78 & M7 III: C        & O-rich non-Mira AGB / SRb        & 27-04-2024 & $600 \times 4$ \\
             &        &      &      &                  &                                  & 12-04-2026 & $300 \times 4$ \\
BQ Gem       & 55383  & 5.00 & 3.80 & M4 III D         & Low-amplitude red giant / SR     & 01-04-2024 & $300 \times 4$ \\
             &        &      &      &                  &                                  & 12-04-2026 & $300 \times 4$ \\
CU Dra       & 121130 & 4.66 & 3.53 & M3.5 III C       & Slow irregular red giant / Lb     & 10-04-2024 & $120 \times 4$ \\
             &        &      &      &                  &                                  & 20-04-2026 & $300 \times 4$ \\
FS Com       & 113866 & 5.61 & 4.05 & M5 III: C        & Low-amplitude pulsating giant     & 26-04-2024 & $30 \times 4$ \\
             &        &      &      &                  &                                  & 20-04-2026 & $300 \times 4$ \\
G Her        & 148783 & 5.01 & 2.72 & M6 III B         & O-rich AGB / SRb                  & 31-03-2024 & $300 \times 4$ \\
             &        &      &      &                  &                                  & 21-04-2026 & $300 \times 4$ \\
LQ Her       & 145713 & 5.70 & 4.39 & M3 III C         & Low-amplitude red giant / SR      & 31-03-2024 & $300 \times 4$ \\
             &        &      &      &                  &                                  & 21-04-2026 & $300 \times 4$ \\
$\omega$ Vir & 101153 & 5.36 & 3.91 & M4.5: III C      & Semiregular/irregular red giant   & 30-04-2024 & $600 \times 4$ \\
             &        &      &      &                  &                                  & 12-04-2026 & $180 \times 4$ \\
$\psi$ Vir   & 112142 & 4.80 & 3.78 & M3 III B         & Slow irregular red giant / Lb     & 10-04-2024 & $300 \times 4$ \\
             &        &      &      &                  &                                  & 12-04-2026 & $180 \times 4$ \\
R Gem        & 53791  & 7.68 & 9.05 & S3.5 - 6.5/6e B   & Mira; S-type AGB                  & 02-04-2024 & $600 \times 4$ \\
             &        &      &      &                  &                                  & 10-04-2026 & $600 \times 4$ \\
RT Vir       & 113285 & 7.41 & 5.09 & M8 III C         & O-rich AGB / SRb                  & 26-04-2024 & $300 \times 4$ \\
             &        &      &      &                  &                                  & 12-04-2026 & $300 \times 4$ \\
RX Boo       & 126327 & 8.60 & 4.37 & M7.5 - M8 C       & O-rich AGB / SRb                  & 28-03-2024 & $600 \times 4$ \\
             &        &      &      &                  &                                  & 20-04-2026 & $300 \times 4$ \\
ST UMa       & 99592  & 6.28 & 5.04 & M4 III C         & Semiregular red giant / SRb       & 26-04-2024 & $300 \times 4$ \\
             &        &      &      &                  &                                  & 10-04-2026 & $300 \times 4$ \\
SW Vir       & 114961 & 6.85 & 4.07 & M7 III: C        & O-rich non-Mira AGB / SRb         & 26-04-2024 & $180 \times 4$ \\
             &        &      &      &                  &                                  & 12-04-2026 & $180 \times 4$ \\
TU CVn       & 112264 & 5.84 & 4.21 & M5 III - IIIa C   & Semiregular red giant / SRb       & 26-04-2024 & $60 \times 4$ \\
             &        &      &      &                  &                                  & 20-04-2026 & $300 \times 4$ \\
TV Gem       & 42475  & 6.56 & 5.17 & M0 - M1.5 Iab C   & Red supergiant / SRc              & 06-03-2024 & $600 \times 4$ \\
             &        &      &      &                  &                                  & 10-04-2026 & $480 \times 4$ \\
U Her        & 148206 & 6.70 & 6.91 & M6.5 - 8+e B      & Mira; O-rich AGB                  & 02-04-2024 & $300 \times 4$ \\
             &        &      &      &                  &                                  & 21-04-2026 & $420 \times 4$ \\
V636 Her     & 151732 & 5.87 & 4.60 & M4.5 III B       & Slow irregular red giant / Lb     & 10-04-2024 & $180 \times 4$ \\
             &        &      &      &                  &                                  & 21-04-2026 & $120 \times 4$ \\
X Her        & 144205 & 6.58 & 3.66 & M6 III C         & O-rich AGB / SRb                  & 31-03-2024 & $300 \times 4$ \\
             &        &      &      &                  &                                  & 21-04-2026 & $60 \times 4$ \\
\enddata

\tablenotetext{}{The $V$ and $G$ band magnitudes and spectral types are adopted from SIMBAD. The $G$ magnitudes correspond to Gaia EDR3 values \citep{brown2021gaia}.}
\end{deluxetable*}


\subsection{Data Reduction}

A complete spectro-polarimetric data set consists of science exposures of the target star obtained at four half-wave-plate (HWP) positions: 0$^\circ$, 22.5$^\circ$, 45$^\circ$, and 67.5$^\circ$, along with calibration frames taken after each science exposure. The raw data from ProtoPol were processed using an in-house data-reduction pipeline developed with custom Python routines based on open-source libraries such as NumPy \citep{harris2020array}, SciPy \citep{virtanen2020scipy}, and Astropy \citep{astropy2013, astropy2018, astropy2022}. The data reduction steps include bias and dark frame subtraction, cosmic-ray removal, order tracing, scattered background light subtraction, sky subtraction, extraction of o- and e- ray intensities for each order, wavelength calibration, and calculation of Stokes parameters from the extracted intensities of the orthogonally polarized rays. The on-sky performance and characterization of the instrument, the development and features of the data analysis pipeline, and the first science results with ProtoPol are described in detail in \cite{maiti2026development}.
\par
Since the present study aims to identify multi-epoch and line-dependent polarization variability, it is important to quantify the polarization uncertainties associated with the ProtoPol observations. Reported polarization variations across emission-line features are generally quite small, typically amounting to a fraction of a percent \citep{oudmaijer1999halpha, vink2005probing}. Consequently, robust statistical detection of such variations requires polarimetric uncertainties to be constrained to roughly 0.1–0.3$\%$. Given that ProtoPol's instrumental polarization has been characterized to be below 0.1$\%$ \citep{maiti2026development}, the instrument is well suited to this type of measurement. Following \citet{patat2006error}, the uncertainties in polarization degree and angle ($\sigma_P = \frac{1}{\sqrt{N/2}(S/N)}$ and $\sigma_{\theta} = \frac{\sigma_P}{2p}$) scale with the signal-to-noise ratio (SNR) achieved at each spectral resolution element, where N denotes the number of half-wave plate (HWP) positions used to derive the Stokes parameters, and  (S/N) is the target SNR. With N=4 for ProtoPol, achieving the desired uncertainty range of $\sim$0.1–0.3$\%$ requires an SNR of approximately 230–700 per spectral resolution element.
\par
Furthermore, since signal strength differs substantially between the peak of an emission feature and its wings, polarization measurements are inherently more precise near the line peak, where SNR is highest, than in the wings or absorption troughs, such as those seen in P-Cygni or double-peaked profiles where SNR is comparatively low \citep{oudmaijer1999halpha, vink2002probing}. To address this, the ProtoPol reduction pipeline implements an adaptive binning scheme that progressively expands the spectral bin width, incorporating additional flux until the target SNR is reached. This approach maintains a uniform polarization uncertainty across bins, though it does so at the expense of spectral resolution.
\par
No correction for instrumental polarization has been applied to the data presented here, since its contribution amounts to a fixed polarization offset, and our analysis focuses on polarization variability across observing epochs rather than on absolute values. For similar reasons, interstellar polarization (ISP) has also not been removed from the data, as it too introduces a constant offset that should remain stable across the different epochs of observation. In addition, ISP exhibits wavelength dependence only across broad spectral ranges \citep{serkowski1974many}, meaning it can reasonably be treated as constant over the comparatively narrow wavelength range spanned by any single emission or absorption feature.
\par
To trace the continuum polarization of the SySt and red giant stars in the sample, the median signal from the central 200 pixels of each echelle order in the red CD setting was used, since the order centers had significantly higher SNR as compared to the order edges, where the signal was low or negligible. Care was taken to avoid any emission/absorption features that may be present in the 200-pixel window. The spectra of the respective stars were relatively flux calibrated using spectrophotometric standard stars observed on nights contemporaneous with the science observations. Full details of the method of relative flux calibration in ProtoPol have been stated in detail in \cite{maiti2026development}. It should be noted that no correction for instrumental polarization has been applied in this work, as the present analysis focuses primarily on relative changes in the polarization signatures across spectral lines and between epochs, rather than on the absolute polarization levels. Therefore, the absolute polarization values may be interpreted with caution. This assumption is justified provided that the instrumental contribution remains stable over the relevant wavelength range and observing epochs.

\section{Results and Discussion}
\label{sec:results-and-discussion}
In this section, we discuss the results obtained from the spectro-polarimetric observations described in Section~\ref{sec:ObservationCampaign}. 
\par
The SySt sample consists of 6 systems observed over multiple epochs with ProtoPol. Symbiotic stars are evolved cool giants paired with hot compact companions in complex circumstellar environments—typically showing strong optical emission lines (H$\alpha$, He I, He II, etc.), with $\sim$50$\%$ exhibiting distinctive $\lambda\lambda$ 6830, 7088 {\AA} features from Raman-scattered O VI $\lambda\lambda$ 1032, 1038 {\AA} in the giant's neutral wind. Spectro-polarimetry is valuable here, as polarization variations across these lines trace scattering geometry, asymmetric winds, outflows, and orbital or epoch-dependent circumstellar changes. In section~\ref{sec:Symbiotic}, we first discuss the overall polarization behavior of the symbiotic sample and then discuss the individual systems separately.
\par
The red giant stars sample consists of 18 systems observed over multiple epochs with ProtoPol. In section~\ref{sec:RedGiants}, we discuss the polarization behavior of the red giant stars with the main focus on continuum polarization rather than strong emission-line polarization, and then discuss the individual systems separately. Since a spherically symmetric unresolved envelope should produce little or no net polarization, the detection of continuum polarization in red giants indicates departures from spherical symmetry in their extended atmospheres or circumstellar envelopes.

\subsection{Symbiotic stars}
\label{sec:Symbiotic}

SySt are ideal targets for optical spectro-polarimetry because their spectra are formed in a complex environment containing a cool giant secondary star, a hot compact companion, ionized gas, and, in many cases, an accretion disc. In particular, the Raman-scattered O~VI emission features at $\lambda\lambda$ 6830, 7088 {\AA} are known to show prominent polarization signatures in several SySt \citep[see e.g.,][and references therein]{schmid1994raman, schild1996spectropolarimetry, harries1996raman, schmid1997spectropolarimetry}. These emission lines are produced when O VI $\lambda\lambda$ 1032, 1038 {\AA} resonance photons, originating close to the hot component, are Raman scattered by neutral hydrogen in the red-giant wind. Several theoretical and numerical studies have significantly advanced our understanding of the underlying scattering physics and circumstellar geometry in symbiotic systems \citep[see e.g.,][and references therein]{schmid1992montecarlo, schmid1995monte, schmid1996simulations, harries1997raman, lee1997profiles}.
\par
An equally significant, but less explored process involves the Raman scattering of far-ultraviolet (far-UV) Ly$\beta$ photons by neutral hydrogen in the wind of the red giant component, giving rise to the broad H$\alpha$ wings observed in many symbiotic stars \citep{lee2000raman, yoo2002polarization}. Broad H$\alpha$ wings are commonly detected in SySt spectra \citep{lee2000raman, chang2018broad}. Hydrodynamic motions within the system have also been proposed as a possible origin, although Raman scattering of far-UV Ly$\beta$ photons by neutral hydrogen has been proposed as a possible cause as well. In this context, spectro-polarimetry has emerged as a powerful diagnostic tool to decouple the two scenarios, since Raman scattering is expected to impart measurable H$\alpha$ polarization values. Indeed, polarized broad H$\alpha$ wings have been detected in systems such as AG Dra and Z And, lending strong support to the Raman-scattering interpretation \citep{ikeda2004polarized}.
\par
Although multi-epoch spectro-polarimetric studies have been conducted in the Raman-scattered O VI features in a few symbiotic systems like AG Dra, V1016 Cyg, SY Mus, etc. \citep{schmid1997spectropolarimetry, schild1996spectropolarimetry, harries1996spectropolarimetric}, to probe the phase-locked variation in polarization with orbital motion of the systems, such studies have never been performed for the Raman-scattered Ly$\beta$ photons. Furthermore, in the few symbiotic systems where H$\alpha$ polarization is reported, such detections were largely restricted to the line wings and exhibited relatively modest polarization amplitudes of $\sim$0.5–0.6$\%$. In contrast, broad polarization profiles with substantially larger amplitudes, in some cases exceeding 10$\%$, have primarily been associated with the Raman-scattered O VI features at $\lambda \lambda$ $6830, 7088 \text{\AA}$ \citep{schmid1994raman, harries1996raman}. As such, the H$\alpha$ polarization detections, as presented in this manuscript, show that multi-epoch observations of such stars is essential to probe the changing polarization, and hence the changing scattering morphologies in such systems.
\par
Raman scattering of Ly$\beta$ photons by neutral hydrogen remains the most compelling mechanism for explaining the observed H$\alpha$ polarization profile; nevertheless, elastic Thomson scattering of H$\alpha$ photons by high-velocity electrons has also been proposed as a possible source of intrinsic H$\alpha$ polarization seen in symbiotics \citep{kim2007Thompson}. H$\alpha$ polarization due to Thomson scattering is seen in several other astrophysical laboratories like Herbig stars, classical Be stars, interacting binaries, etc \citep{harrington2007spectropolarimetry, oudmaijer1999halpha, brown1978polarisation, rudy1978polarimetric}. If such is the scenario, the broad H$\alpha$ wings may arise through electron scattering, wherein H$\alpha$ photons are scattered by rapidly moving electrons, thereby producing significant line broadening and intrinsic polarization \citep{kim2007Thompson, chang2018broad}.
The Thomson-scattering optical depths and electron temperatures of several symbiotic systems such as AG Dra, Z And, and V1016 Cyg have previously been estimated from analysis of the broadened wings of O VI and He II emission lines in the UV spectra \citep{sekeravs2012electron}. However, observational evidence conclusively supporting electron scattering of H$\alpha$ photons as the dominant origin of broad wings remains limited. One notable example is the spectro-polarimetric study of the symbiotic star BI Crucis, in which electron scattering was proposed as the mechanism responsible for the polarized H$\alpha$ wings \citep{harries1996accretion}.
\par
Apart from the polarization detection in the Raman scattered emission features, SySt, due to the presence of the red giant companion, are known to show strong and variable continuum polarization. This polarization originates due to starlight scattering off circumstellar dust grains of the giant companion \citep{bieging2006optical}. The magnitude of the polarization depends on the wavelength, grain size, shape, and composition. Imaging polarimetric studies of a symbiotic star sample have been done by \cite{schulte1990polarimetric}, showing their continuum polarization properties to be similar to AGB/post-AGB stars. \citep{brandi2000linear} conducted multi-epoch imaging polarimetric studies of a symbiotic star sample and showed the dynamic polarization nature of the sources. A higher-resolution multi-epoch spectro-polarimetric survey of the continuum polarization of such stars would thus be beneficial for the community to study the temporal variation of continuum polarization.   
\par
\begin{figure}[]
  \centering

  \begin{subfigure}[b]{0.49\textwidth}
    \includegraphics[width=\textwidth]{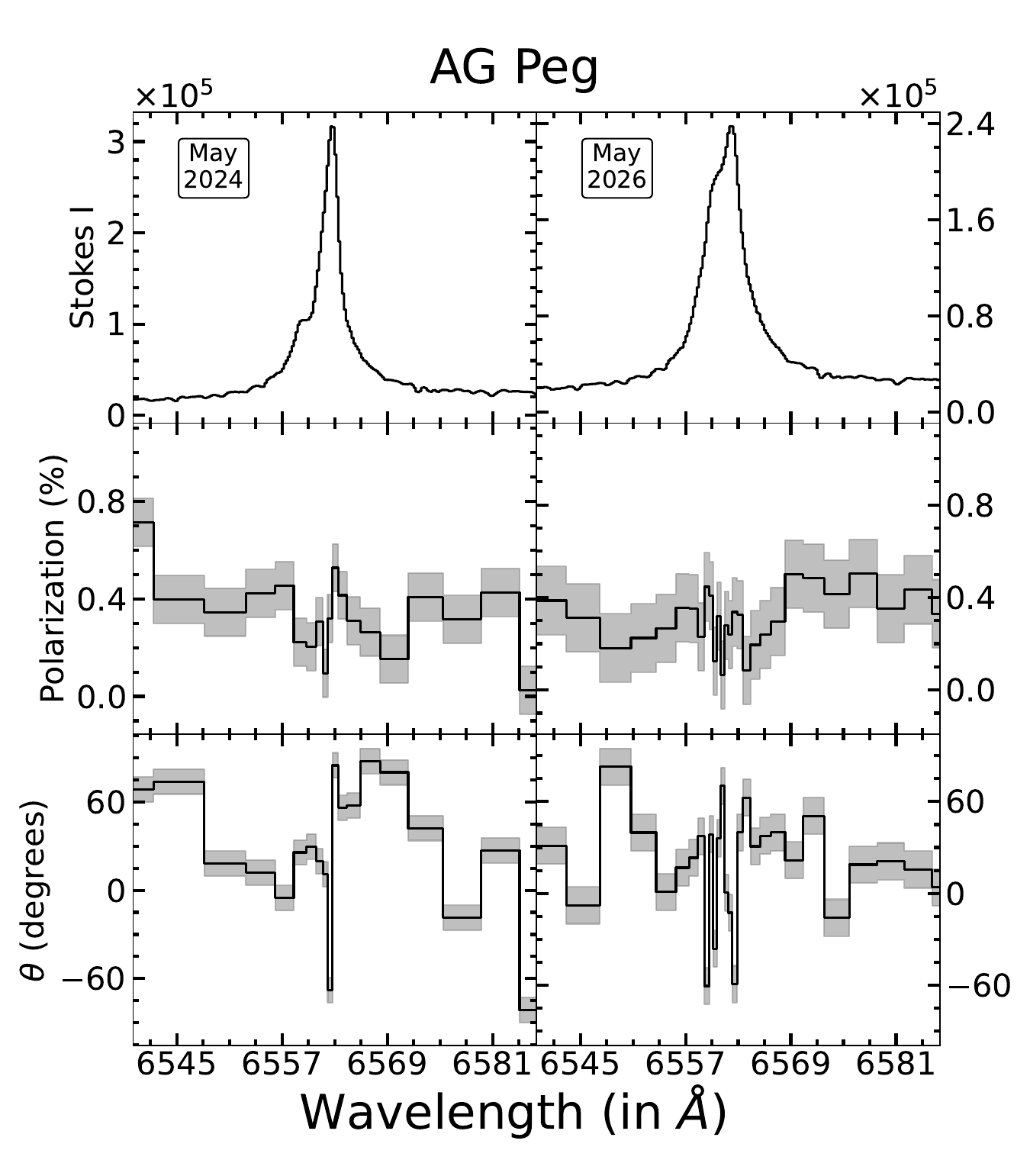}

    \label{fig:1}
  \end{subfigure}
  \begin{subfigure}[b]{0.49\textwidth}
    \includegraphics[width=\textwidth]{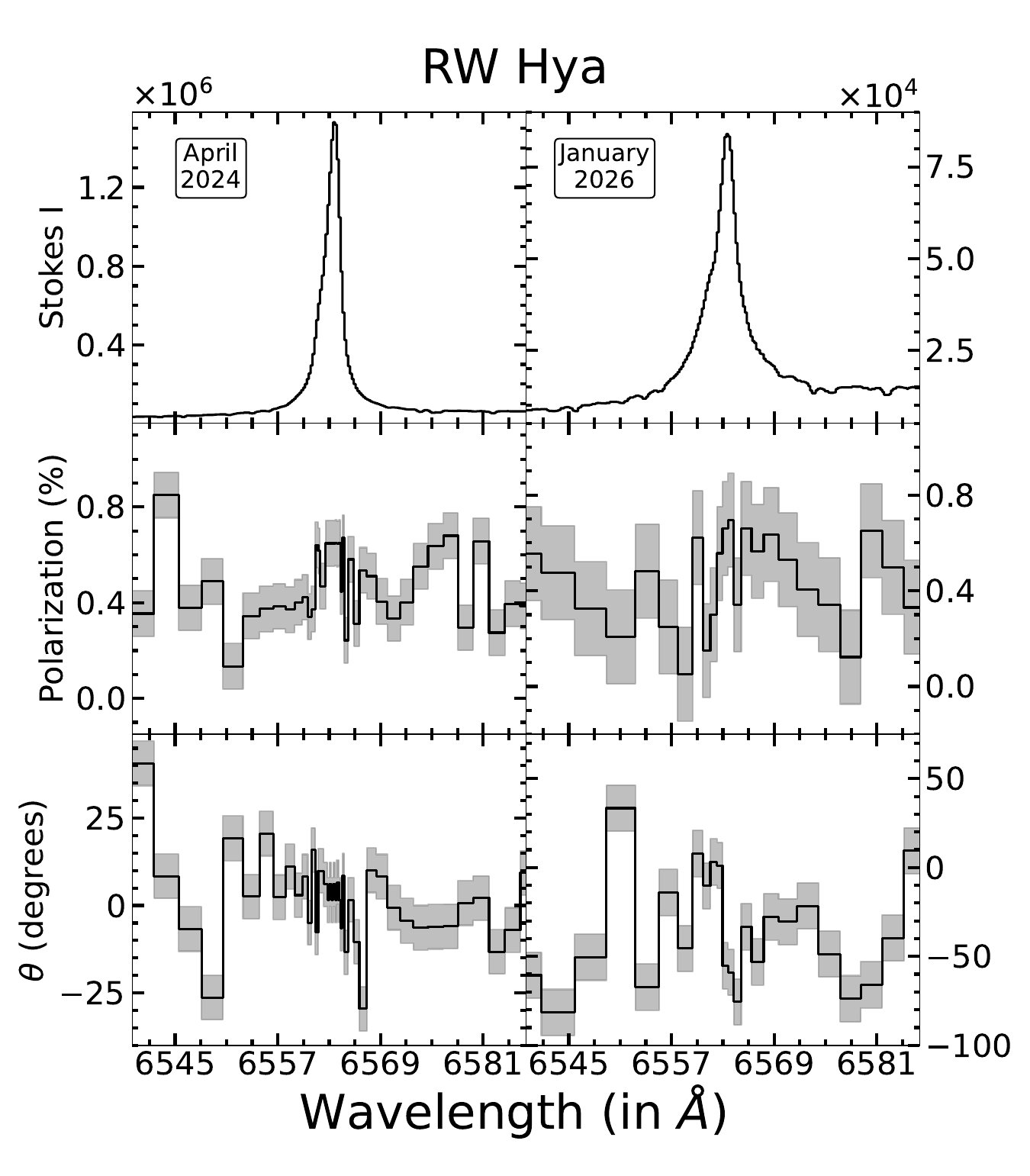}

    \label{fig:2}
  \end{subfigure}

  \vspace{0.1cm}

  \begin{subfigure}[b]{0.49\textwidth}
    \includegraphics[width=\textwidth]{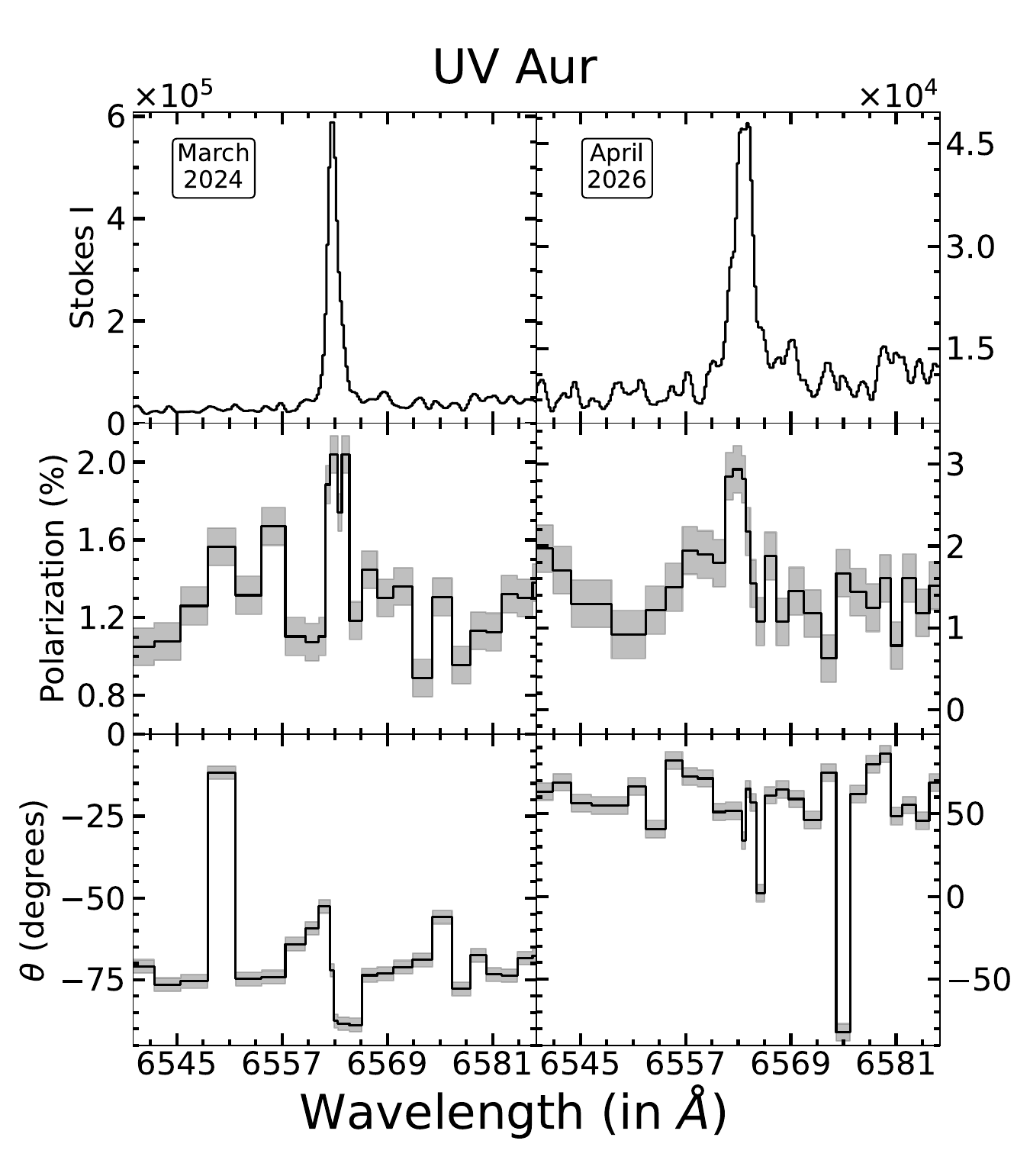}

    \label{fig:3}
  \end{subfigure}

\caption{Polarization spectra for symbiotic stars AG Peg, RW Hya, and UV Aur. For each star, the Stokes I spectrum is shown in the top panel of the triplot, the degree of polarization is shown in the middle panel, and the angle of polarization is shown in the bottom panel. The left panel of each star shows the first epoch data, while the right panel shows the second epoch data. The data have been dynamically binned to achieve a constant polarization accuracy over each bin. The corresponding errors in the detection of p and $\theta$ are shown with a gray shaded strip.}
  \label{fig-symbiotic_1}
  
\end{figure}
\begin{figure*}
  \centering

  \begin{subfigure}[b]{0.49\textwidth}
    \includegraphics[width=\textwidth]{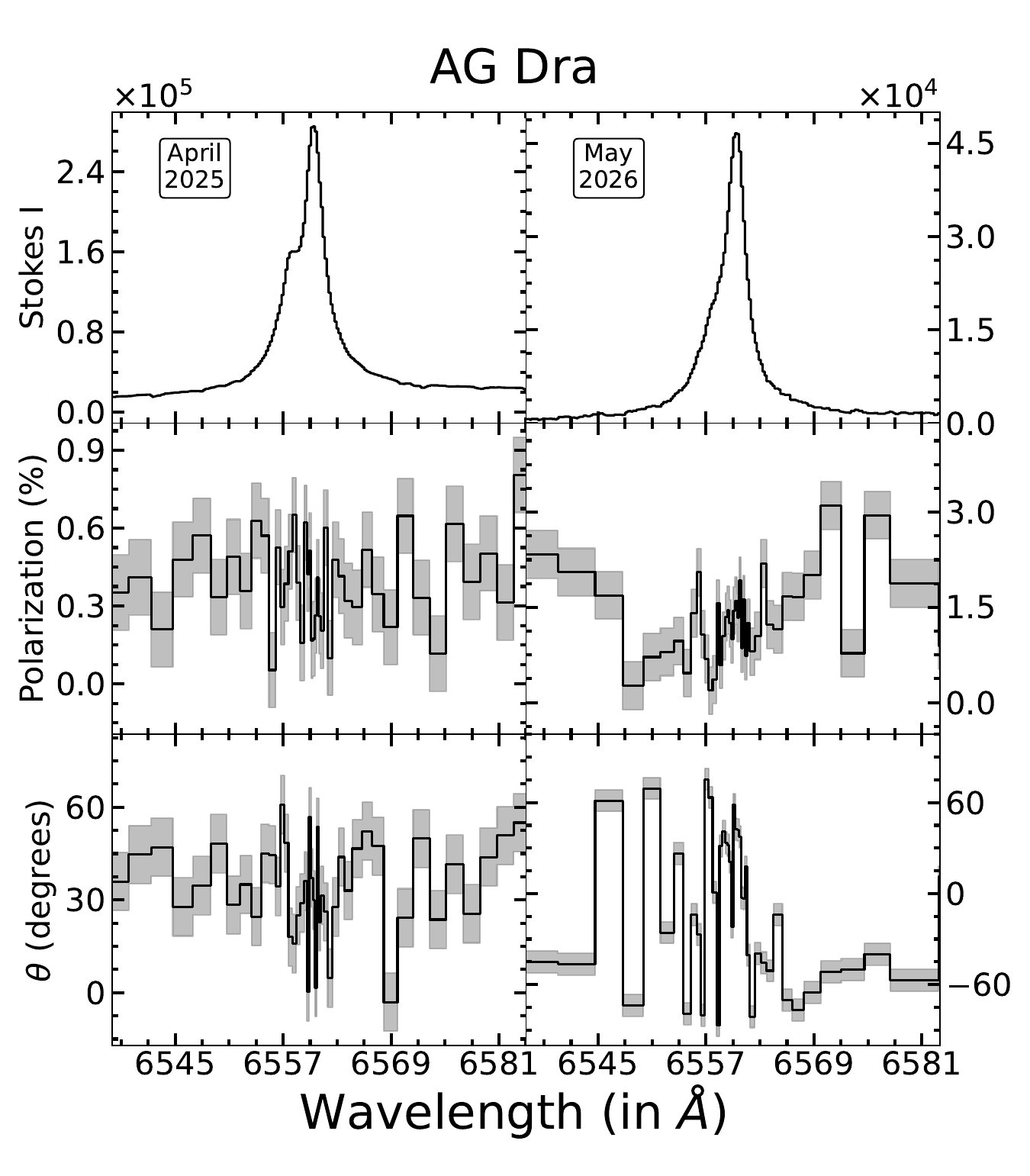}

    \label{fig:1}
  \end{subfigure}
  \begin{subfigure}[b]{0.49\textwidth}
    \includegraphics[width=\textwidth]{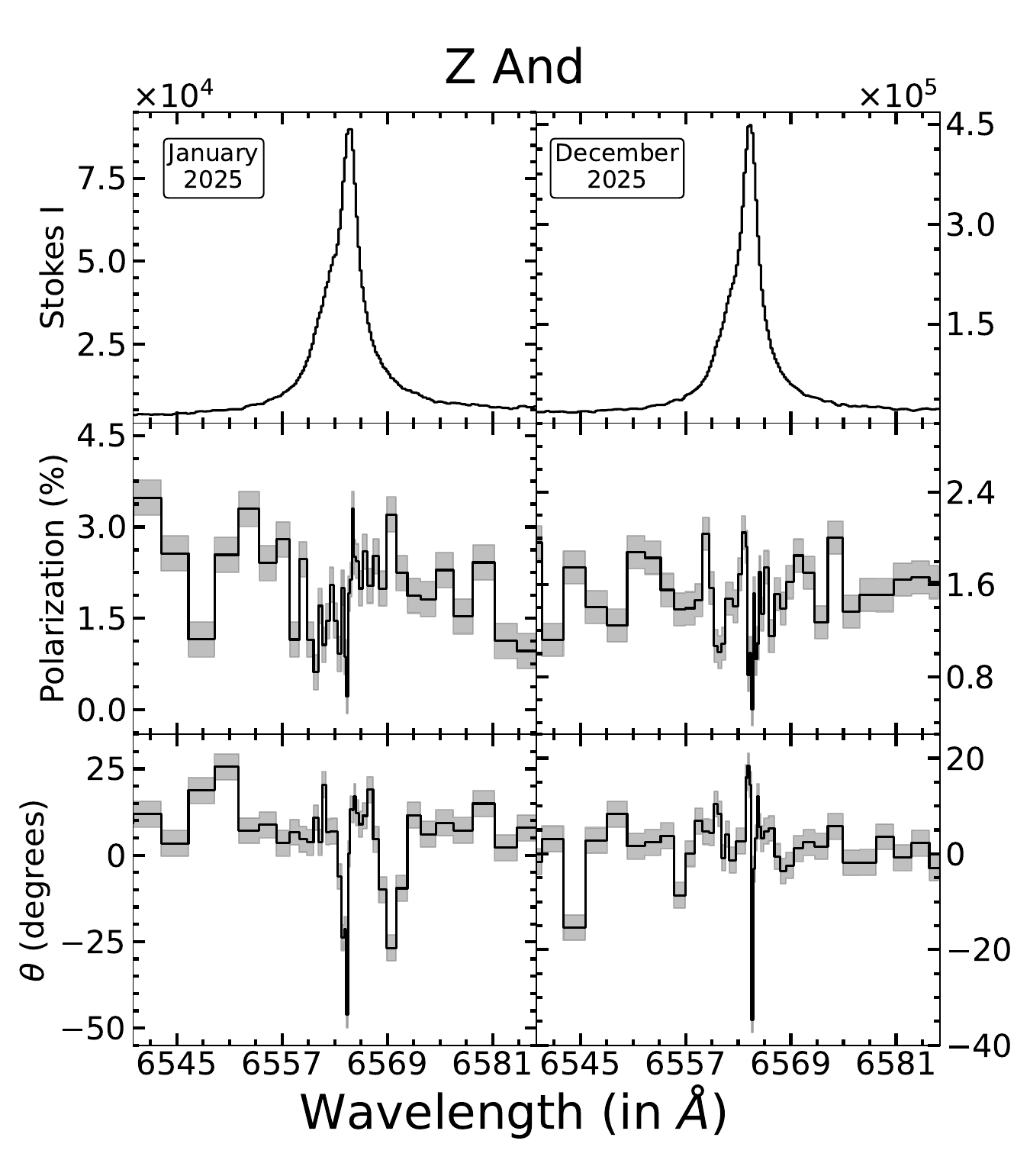}

    \label{fig:2}
  \end{subfigure}
  
  \caption{Same as Figure~\ref{fig-symbiotic_1}, but for symbiotic star AG Dra (left) and Z And (right).}
  \label{Fig-symbiotic_2}
\end{figure*}
\begin{figure}[]
  \centering

  \begin{subfigure}[b]{0.49\textwidth}
    \includegraphics[width=\textwidth]{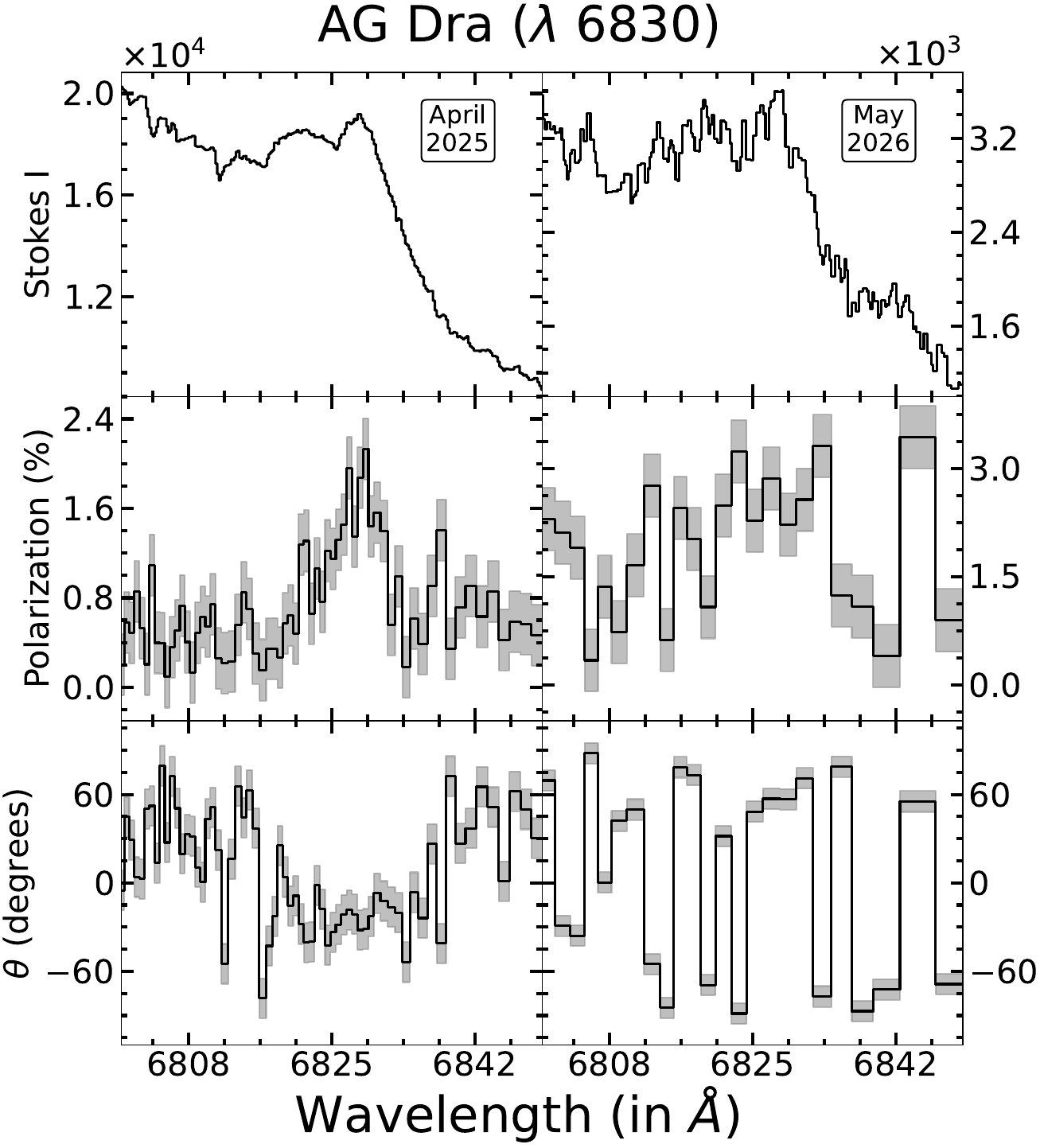}

    \label{fig:1}
  \end{subfigure}
  \begin{subfigure}[b]{0.49\textwidth}
    \includegraphics[width=\textwidth]{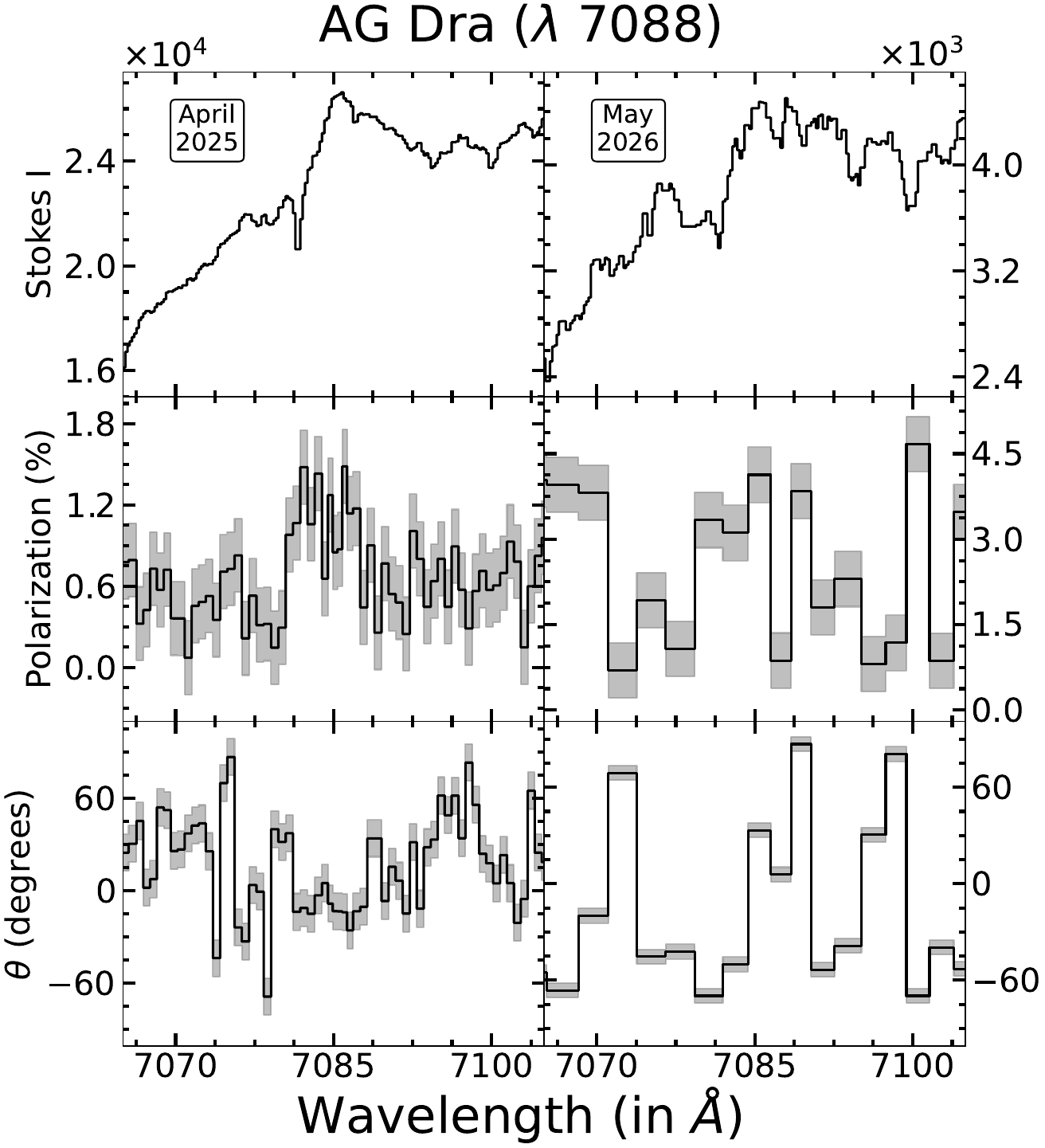}

    \label{fig:2}
  \end{subfigure}

  \vspace{0.1cm}

  \begin{subfigure}[b]{0.49\textwidth}
    \includegraphics[width=\textwidth]{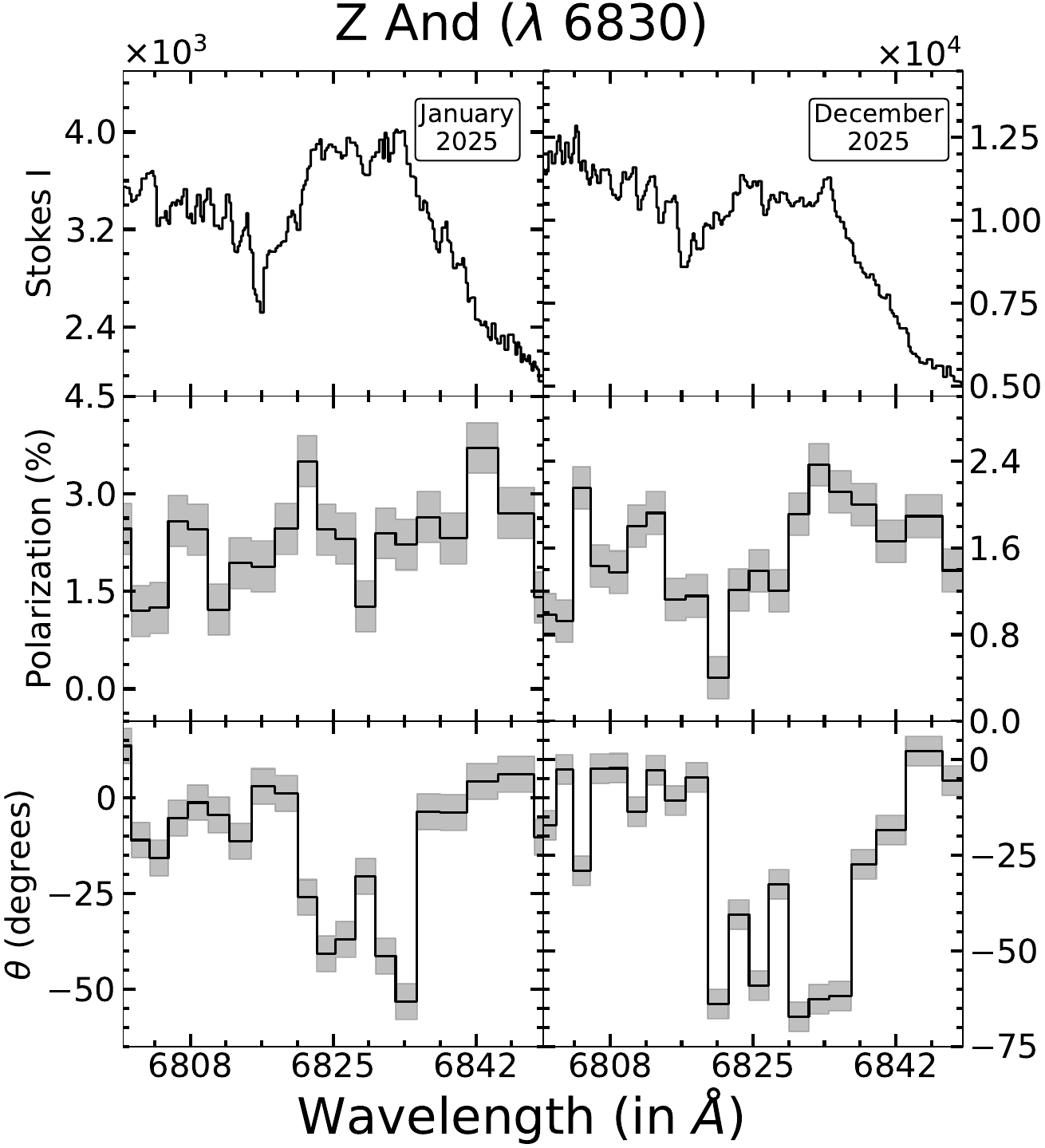}

    \label{fig:3}
  \end{subfigure}
  \begin{subfigure}[b]{0.49\textwidth}
    \includegraphics[width=\textwidth]{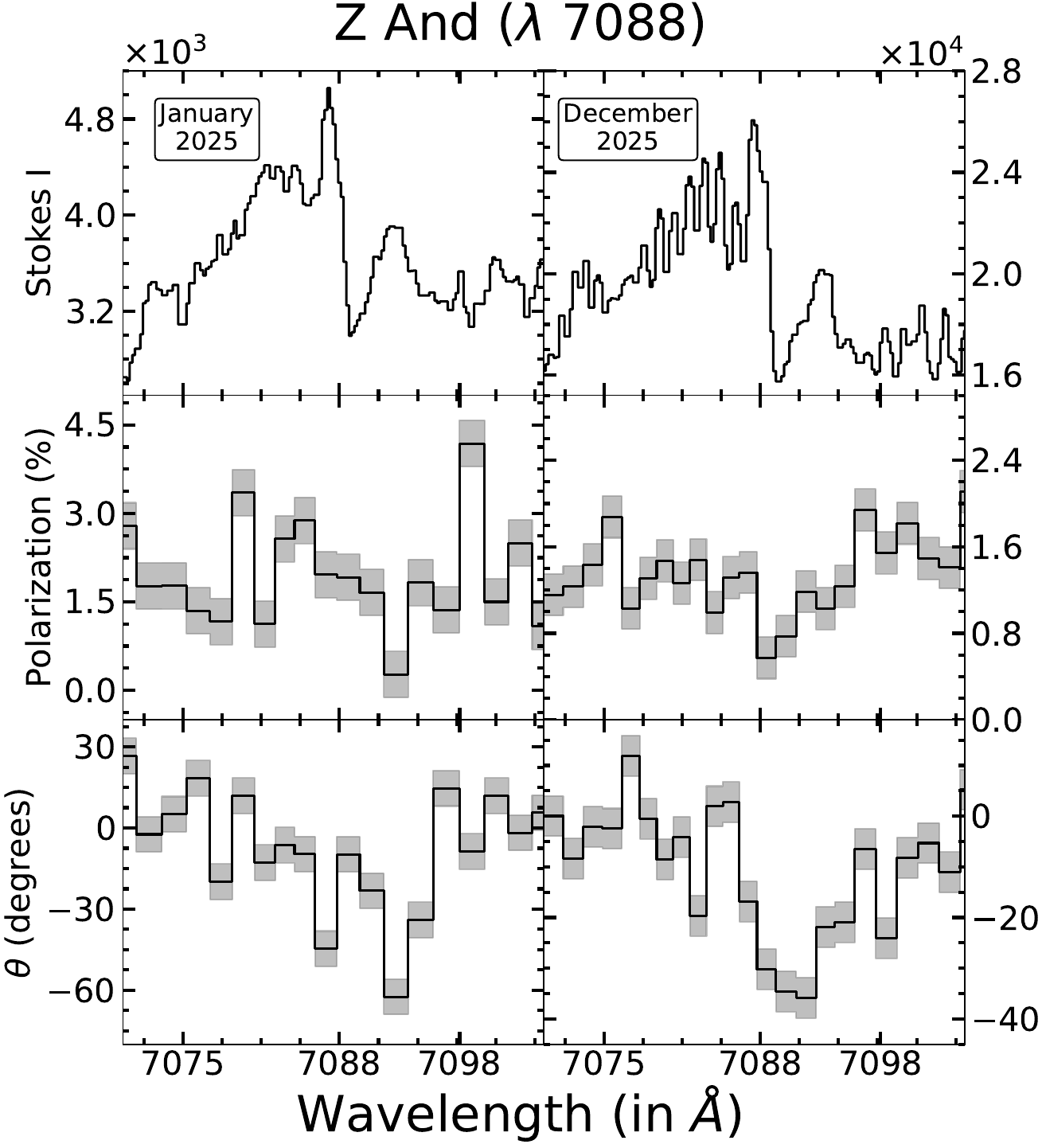}

    \label{fig:4}
  \end{subfigure}

  \caption{Same as Figure~\ref{fig-symbiotic_1}, but for symbiotic star AG Dra (top panel) and Z And (bottom panel) for Raman scattered $\lambda \lambda$ $6830, 7088 \text{\AA}$ emission features.}
  \label{fig-symbiotic_3}
  
\end{figure}
\begin{figure*}
  \centering
  \includegraphics[width=\textwidth]{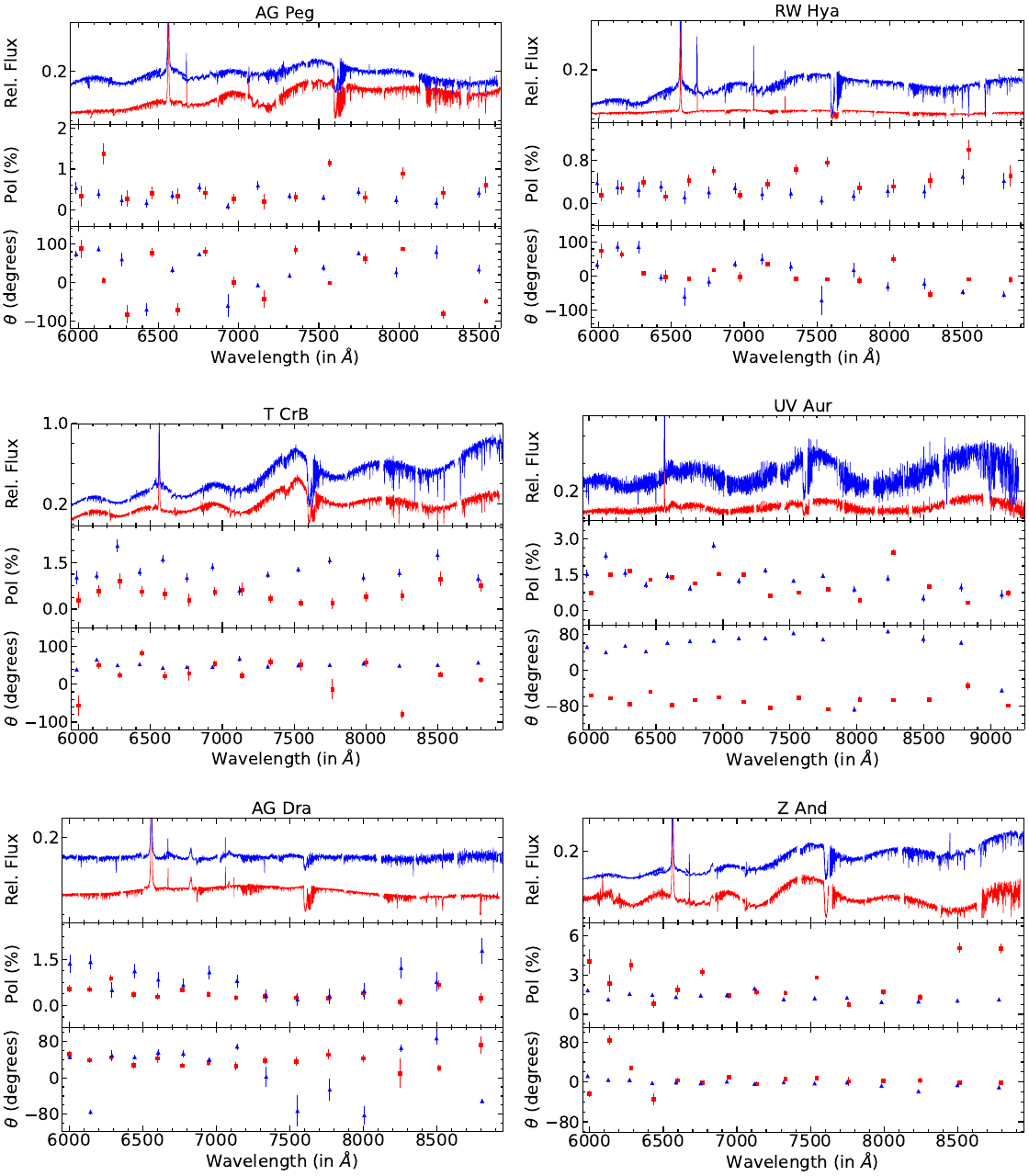}  
  \caption{Continuum polarization of the sample of symbiotic stars, namely AG Peg, RW Hya, T CrB, UV Aur, AG Dra, and Z And. For each star, the top panel shows the relative flux calibrated spectra of the star; the middle and the bottom panels show the degree and angle of polarization calculated from the median signal of the central 200 pixels of each order. The first and second epochs of observations are shown in red squares and blue triangles, respectively. The relative flux calibrated spectra of the second epoch have been given a manual shift of 1 unit for better clarity.}
  \label{Fig-symbiotic_cont}
\end{figure*}

The multi-epoch H$\alpha$ polarization spectra are shown in Figures~\ref{fig-symbiotic_1} and \ref{Fig-symbiotic_2}, while the polarization behavior of the Raman-scattered $\lambda\lambda$6830, 7088 {\AA} features is shown in Figure~\ref{fig-symbiotic_3}. The continuum polarization measurements are shown in Figure~\ref{Fig-symbiotic_cont}. In the present sample, only a few systems show clear polarization changes across H$\alpha$, whereas variable polarization is detected across the Raman-scattered O~VI features whenever these features are present with sufficient strength. The continuum polarization is also variable in several objects, indicating changes in the scattering geometry or relative contributions from intrinsic and interstellar polarization. Since the data have not been corrected for interstellar polarization, the continuum polarization values should be interpreted as the combined contribution of intrinsic polarization from asymmetric circumstellar material and interstellar polarization along the line of sight.

\paragraph{AG Peg}
It is a symbiotic binary with an MIII giant and a white dwarf (WD) in an 818-day orbit \citep{fekel2000infrared}. The system underwent a long-lasting nova-like outburst beginning around the mid-nineteenth century, followed by a slow decline over more than a century \citep{kenyon1993evolution}. AG Per shows a faint Raman-scattered $\lambda$6830 {\AA} emission feature, which became prominent during its 2015 outburst \citep{skopal2017new}. ORFEUS UV spectroscopy also showed the presence of the O VI $\lambda \lambda$ 1032, 1038 {\AA} resonance doublet lines, with the line strength varied significantly across different epochs \citep{schmid1999orfeus}. In the observations from ProtoPol, however, the Raman-scattered feature is very weak in the first epoch and is not detected in the second epoch. Polarimetric observations of AG Peg have been reported in several earlier studies \citep[see e.g.,][]{serkowski1974many, schulte1990polarimetric, brandi2000linear}, showing variability in polarization both with wavelength and time. The reported polarization amplitudes are generally small, reaching $\sim$ 1\% in the $B$ band in \cite{serkowski1974many}, and only $\sim$ 0.2\% in the $U$ band in \cite{schulte1990polarimetric}. Multi-epoch imaging polarimetry over nearly three years by \cite{brandi2000linear} showed significant changes in the polarization amplitude, along with rotations of the polarization angle. Similar rotations of the polarization angle have also been discussed in earlier work \citep{coyne1988polarized}. The continuum-polarization measurements from ProtoPol show comparable polarization levels in both epochs, with a mean value of approximately $\sim$ 0.4 \%. A rapid wavelength-dependent rotation of the polarization angle is also observed. No previous spectro-polarimetric observations of AG Peg could be found in the literature, either for the Raman-scattered $\lambda\lambda$6830, 7088 {\AA} features or for H$\alpha$. Since the Raman-scattered $\lambda\lambda$6830, 7088 {\AA} features are weak or absent in the ProtoPol data, we present the multi-epoch H$\alpha$ polarization measurements. No clear polarization change is detected across the H$\alpha$ profile in either epoch, with the polarization remaining nearly constant at $\sim$0.4\% across the line.
\par
\paragraph{RW Hya} 
It is an eclipsing binary system with an orbital period of 370 days \citep{schild1996high}. Its symbiotic nebulosity lies very close to the red giant and may be considered part of the extended red giant atmosphere. RW Hya is a relatively dormant symbiotic system, with no recorded outburst to date; the observed light-curve variations are mainly associated with orbital motion. The width of the H$\alpha$ profile also varies significantly between eclipse and out-of-eclipse phases \citep{schild1996high}. Based on IUE UV spectral analysis, \cite{sion2002iue} suggested that the hot component of RW Hya is a young, accretion-heated, nuclear shell-burning WD on its final cooling track. Observations with ProtoPol of RW Hya show a nearly constant continuum polarization of $\sim$0.4\% in the second epoch, whereas the first epoch shows an increase in polarization toward redder wavelengths. A similarly low and nearly constant polarization was reported by \cite{schulte1990polarimetric}, who noted that the polarization properties were consistent with an interstellar origin. However, they also reported significant continuum-polarization variations over a period of four months. RW Hya does not show Raman-scattered $\lambda\lambda$6830, 7088 {\AA} emission features in its spectra. No previous spectro-polarimetric study of this target could be found in the literature. The multi-epoch ProtoPol spectro-polarimetric observations of the H$\alpha$ profile, separated by nearly 21 months, do not show any noticeable polarization change across the emission-line profile. A higher-cadence spectro-polarimetric monitoring campaign over different orbital phases, from conjunction to quadrature, would be useful for testing possible phase-dependent polarization variability across the emission-line profiles.
\par
\paragraph{UV Aur:} 
It is a carbon Mira with a photometric period of 394 days \citep{herbig2009carbon}. It was classified as a symbiotic star from observations of [O III] and [Ne III] emission lines in their optical spectra \citep{sanford1949phases, sanford1950variations}, and later by the detection of [Fe VII] \citep{seal1988symbiotic}. Its Balmer emission lines are heavily muted by overlying C-type absorption bands. Imaging polarization observations of the star by \cite{schulte1990polarimetric} showed a decrease in polarization from optical to infrared, with peak polarization $\sim$1.5$\%$ at B-band center. \cite{khudyakova1985polarimetry}, through 11-year polarimetric observations, showed that the polarization detected in UV Aur is partly intrinsic, and that the polarization varied periodically with Mira pulsations. The correlation between the polarization in the Mira pulsation phase suggests the presence of some photospheric scattering. Observations with ProtoPol over two epochs separated by over 2 years show a similar polarization trend as reported in the literature, with the polarization steadily decreasing from bluer to redder wavelengths. A similar trend is observed for both epochs of observation in the degree of polarization, although an almost 120-degree rotation in the polarization angle is seen between the two epochs. The continuum polarization spectrum of UV Aur can be well described by the $\lambda^{-1}$ law, which strongly indicates Mie scattering as the cause of the polarization \cite{schulte1990polarimetric}. More interestingly, the H$\alpha$ profile of the star shows an increase in the polarization across the profile, with peak polarization increasing from $\sim$0.5 to 1.0$\%$ above continuum polarization from the first to second epoch. The width of the polarization feature is comparable to the width of the H$\alpha$ emission line itself. Both Raman scattering of Ly$\beta$ photons in the neutral hydrogen wind of the Mira, or Thompson scattering of the H$\alpha$ photons themselves in the electron wind, could be probable causes for the observed polarization. Further spectro-polarimetric follow-up observations of the star have been planned to explore the H$\alpha$ polarization variability. 


\paragraph{T CrB:} It is a famous symbiotic recurrent nova with a recurrence period of 80 years, with previous nova outbursts observed in 1866 and 1946, since when the star has been in quiescence \citep{kenyon1986symbiotic}. In the quiescent phase, the star has been accreting at a very slow rate, showing very few emission lines (typically H$\alpha$) in the optical spectra, which is otherwise dominated by broad absorption bands of the MIII red giant companion \citep{munari2025t}. A drastic increase in the mass transfer rate was observed, characterized by the appearance of several emission lines in the optical spectra, similar to the enhanced mass transfer event about 8 years before the 1946 outburst \citep{munari20162015}, thus proposing the next outburst event to occur around 2025-2026. This has led to T CrB being one of the most intensely monitored systems in the world, leading to a wealth of multi-wavelength observations, which would enhance our understanding of symbiotic novae in general. With an orbital period of $\sim$227 days, \cite{munari2025t} presented a detailed catalog of the evolution of various emission lines in T CrB spectra with phase. The same is evident from several observations from ProtoPol, taken for a period of over 16 months (2 orbital cycles), with the H$\alpha$ emission profile varying from single-peaked to double-peaked, with varying strengths of the blue- and red-shifted peaks. No Raman scattered  $\lambda \lambda$ $6830, 7088 \text{\AA}$ emission features were detected in any of the observation epochs. The spectro-polarimetric results show no significant polarization change across the H$\alpha$ profile as compared to the continuum. However, the average continuum polarization values have increased from $\sim$0.5$\%$ in early 2025 to $\sim$1.1$\%$ this year. \cite{nikolov2022interstellar}, through optical spectro-polarimetric observations conducted from 2018-2021, predicted a polarization maximum of 0.46$\%$ at 5200$\AA$, with the author claiming that T CrB had no intrinsic polarization and that the measured polarization is due to the line-of-sight interstellar dust. Our polarization measurements in the first epoch are very close to the reported values, but the second epoch of observations indicates that T CrB has developed significant ($\sim$1.0$\%$) intrinsic polarization. The system would be thoroughly monitored spectro-polarimetrically, through its expected outburst, with ProtoPol. Any significant changes in the polarization profile would be promptly reported in future work.

\begin{figure*}
  \centering
  \includegraphics[width=\textwidth]{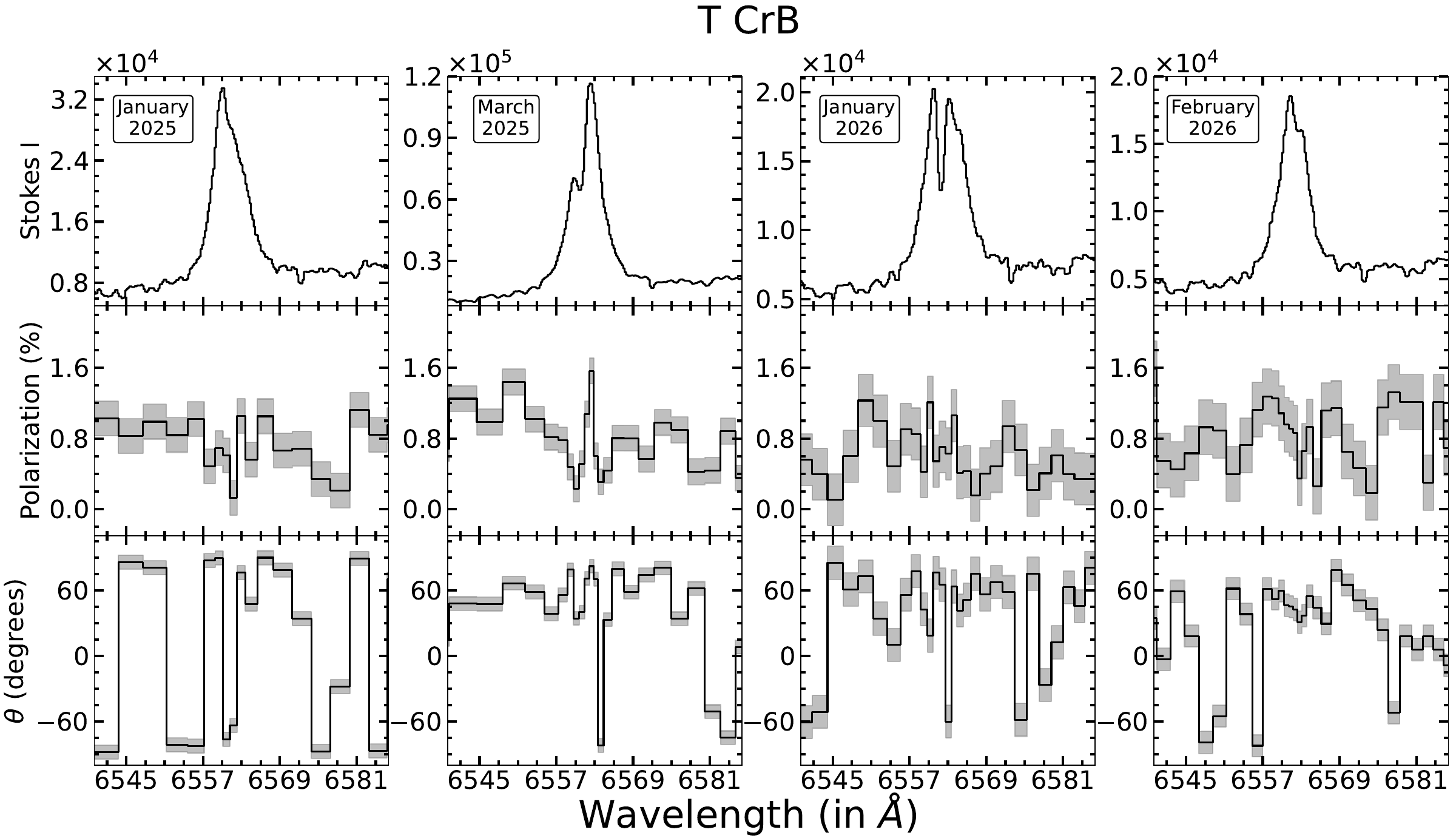}  
  \caption{Same as Figure~\ref{fig-symbiotic_1}, but for symbiotic star T CrB over four epochs of observation.}
  \label{Fig-symbiotic_4}
\end{figure*}

\paragraph{AG Dra:} It is one of the most spectro-polarimetrically well-studied symbiotic star sources in the literature, due to being relatively bright and showing the presence of Raman scattered $\lambda \lambda$ $6830, 7088 \text{\AA}$ emission features in its optical spectra. Extensive photometric and spectroscopic work has been done on the source to constrain several of the system parameters \citep{iijima1987spectroscopic, mikolajewska1995evolution}, like the temperature of the hot white dwarf (T$\approx$120000 K) and properties of the accompanying K giant, while also establishing the 549-day period of the system \citep{fekel2000infrared}. Spectro-polarimetric observations of the Raman scattered $\lambda \lambda$ $6830, 7088 \text{\AA}$ emission features were first conducted by \cite{schmid1994raman} in a spectro-polarimetric survey of Raman features in a sample of symbiotic stars, where he reported polarization $\sim$1$\%$ and $\sim$0.5$\%$ above continuum for the $\lambda$6830 and $\lambda$ 7088 $\text{\AA}$ emission features, respectively. \cite{schmid1997spectropolarimetry} conducted a phase-resolved spectro-polarimetric campaign of the star over almost 4 years (including the 1994 outburst phase of the star), to show the near phase-locked variation of the polarization angle and the temporal variability of the degree of polarization of the Raman scattered $\lambda$6830 feature. He proposed the system inclination to be around 120 degrees and the orbit orientation around 150 degrees from his multi-epoch survey. Later, \cite{ikeda2004polarized} detected a faint polarization signature $\sim$0.5$\%$ in the H$\alpha$ wings of AG Dra, proposing Raman scattering of Ly$\beta$ photons in the neutral hydrogen wind of the giant as the probable cause. Observations with ProtoPol, taken over two epochs separated by more than 2 years, confirm the above-stated observations. A variable polarization in the Raman-scattered $\lambda \lambda$ $6830, 7088 \text{\AA}$ emission features was detected in ProtoPol observations; a clear increase in polarization was seen in the first epoch for both emission features; however, almost no polarization change was observed during the second epoch. No visible polarization change could be noticed across the H$\alpha$ emission for either epoch. A slight depolarization could be seen in the second epoch, but that could be an artifact of low SNR in the neighboring continuum. The continuum polarization remains almost constant ($\sim$0.5$\%$) across the entire wavelength range during the first epoch of observation; however, in the second epoch, an increase in polarization is noticed, both in the bluer and redder wavelengths.  

\paragraph{Z And} 
It is a prototypical SySt with a well-established period of 759 days \citep{fekel2000infrared}. Like AG Dra, it is one of the most extensively studied symbiotic systems in spectro-polarimetry. Polarization across the Raman-scattered $\lambda\lambda$ 6830, 7088 {\AA} emission features in Z And was first reported by \cite{schmid1994raman}. Later, \cite{schmid1997polarimetric} conducted an extensive multi-epoch spectro-polarimetric survey of the source over one orbital cycle and showed phase-locked variations in both the degree and angle of polarization, with the integrated line polarization varying between 6 and 9\%. From this survey, the system inclination and orbital orientation were estimated to be 47$^\circ$ and 72$^\circ$, respectively. In addition, \cite{ikeda2004polarized} detected a faint polarization signature, $<0.5\%$, in the H$\alpha$ wings. Using low-resolution multi-epoch spectro-polarimetry, \cite{isogai2010orbital} reported continuum polarization values in the range of 1.0 - 1.5\%, peaking around 4500 {\AA} and decreasing from blue to red wavelengths. Observations from ProtoPol show continuum polarization at the level of $\sim$ 1 - 2\%. The continuum polarization values determined for the first epoch have more scatter, but that is probably due to poor SNR in some of the echelle orders during that epoch. No noticeable polarization across the H$\alpha$ line was observed for either of the epochs of observation. Similarly, no significant change in the degree of polarization is seen across the Raman-scattered $\lambda\lambda$ 6830, 7088 {\AA} emission features in either epoch; however, a distinct rotation of the polarization angle is observed across both Raman features in both epochs.

\subsection{Red giant stars}\label{sec:RedGiants}
The extended atmospheres and dusty circumstellar envelopes of giant stars provide natural scattering regions, making them significant candidates for spectro-polarimetric studies \citep[][and references therein]{bieging2006optical, 1994AJTrammell, 2018A&AHofner}. The detection of intrinsic continuum polarization suggests a deviation from spherical symmetry in the stellar atmosphere or circumstellar environment because the polarization vectors generated by scattering cancel out in an unresolved, spherically symmetric envelope, leaving little or no net linear polarization \citep{1977A&ABrown, brown1978polarisation, 2009A&AIgnace}. These asymmetries may arise from clumpy or episodic dust formation, pulsation-driven shocks, asymmetric molecular layers, or bipolar/axisymmetric mass-loss structures, etc., \citep{bieging2006optical, 2016A&AOhnaka, schmid1999orfeus}. Optical spectro-polarimetric observations of AGB and post-AGB stars have shown that polarization is often wavelength-dependent, indicating contributions from dust scattering, molecular opacity, and changing circumstellar geometry \citep{bieging2006optical, 1994AJTrammell}. High-angular-resolution polarimetric observations have further shown that dust can form close to the stellar photosphere, within only a few stellar radii, and that scattering by large, relatively transparent grains may contribute to wind driving in oxygen-rich AGB stars \citep{2012NaturNorris, 2018A&AHofner}. Therefore, continuum polarization offers a sensitive diagnostic of mass-loss geometry and weak envelope asymmetry in red giants, even in objects whose circumstellar envelopes are too faint or too compact to be directly resolved. In the following subsections, we discuss the spectro-polarimetric observations of the 18 red giant stars observed with ProtoPol.

\paragraph{BK Vir } 
It is a late-type M giant and a low-amplitude semi-regular variable, commonly classified as an SRb-type long-period variable \citep{2017ARepSamus}. Such SRb variables are generally associated with evolved red giants on or close to the AGB and differ from Mira variables by having smaller amplitudes and less strictly periodic light variations \citep{2003A&AKnapp}. BK Vir is therefore a useful representative of the non-Mira, oxygen-rich AGB population, for which the onset of mass loss and atmospheric extension are still not fully understood  \citep{2012A&AOhnaka, 2002A&AOlofsson}. High-angular-resolution VLTI/AMBER observations of BK Vir have spatially resolved its CO first-overtone line-forming region and revealed that the molecular atmosphere is more extended than predicted by hydrostatic photospheric models, with evidence for asymmetry and possible temporal variability in the CO-forming layers \citep{2012A&AOhnaka}. In the context of optical spectro-polarimetry, BK Vir is therefore a particularly useful target for probing weak asymmetries in the molecular and dusty environment of low-amplitude AGB stars. The ProtoPol observations of BK Vir	show low-to-moderate continuum polarization, mostly below $\sim$ 1\%. The two epochs are broadly similar in $P$, although the polarization angle differs between epochs. If the observed polarization is intrinsic to the source, a wavelength-dependent polarization signature can be interpreted as evidence for scattering by an asymmetric extended atmosphere, molecular opacity effects, or weak circumstellar dust, while a nearly wavelength-independent component may indicate interstellar polarization or a weakly varying instrumental/continuum offset \citep{2009A&AIgnace, maiti2026development}.
\par
\paragraph{BQ Gem} 
BQ Gem, also known as 51 Gem, is a bright late-type M giant, usually classified as a semi-regular variable with spectral type around M4 III \citep{1964ApJSSmak, 1977A&ABrown}. It belongs to the low-amplitude red-giant/early-AGB population rather than the classical large-amplitude Mira variables. CO observations of nearby short-period AGB candidates did not detect molecular emission from BQ Gem, suggesting that its present-day circumstellar envelope and mass-loss activity are relatively weak compared with dustier AGB stars \citep{2018MNRASMcdonald}. The ProtoPol observations of BQ Gem show weak polarization, generally $<$ 0.7\%, with no strong wavelength-dependent trend. The first epoch appears slightly more scattered than the second epoch. The low $P$ and variable $\theta$ suggest that any intrinsic polarization is weak. Therefore, the weak continuum polarization detected with ProtoPol would most likely trace weak atmospheric asymmetry, surface inhomogeneity, molecular-layer scattering, interstellar polarization, or low-level circumstellar dust, rather than a dense dusty envelope \citep{1977A&ABrown, 2009A&AIgnace, maiti2026development}.

\paragraph{CU Dra} 
CU Dra, also known as 10 Dra, is a bright late-type M giant with spectral type around M3.5 III and is generally classified as a slow irregular variable of Lb type \citep{2017ARepSamus}. Such low-amplitude irregular red giants are less extreme than Mira variables and usually represent a weaker pulsation and mass-loss regime \citep{2003A&AKnapp}. CU Dra is therefore useful as a comparison object for testing whether relatively low-amplitude evolved red giants show detectable intrinsic continuum polarization. The ProtoPol observations of CU Dra show weak continuum polarization, mostly $<$ 0.5\%, with modest epoch-to-epoch differences. The polarization angle is scattered, particularly where $P$ is low. Overall, CU Dra appears to have one of the weaker polarization signatures in the sample.

\paragraph{FS Com} 
It is a late-type M giant and a small-amplitude pulsating red giant, generally treated as a semi-regular or irregular variable rather than a classical Mira star. Long-term photometric studies of nearby M giants show that such stars often display multiple short periods and low-amplitude variability, indicating pulsation in the upper-RGB or early-AGB regime \citep{1999A&AKiss, 2009MNRASTabur}. The ProtoPol observations of FS Com show moderate polarization with noticeable epoch-to-epoch changes, especially at shorter wavelengths where the first epoch reaches higher $P$. The polarization angle is variable, suggesting a weak but possibly changing intrinsic scattering component. Since FS Com is not known as a strongly dust-enshrouded or high mass-loss AGB star, this continuum polarization detected with ProtoPol would likely trace weak atmospheric asymmetry, molecular-layer scattering, surface inhomogeneity, interstellar polarization, or low-level circumstellar material \citep{1977A&ABrown}.

\paragraph{G Her} 
It is a cool oxygen-rich M giant and a semi-regular variable, commonly classified as an SRb-type AGB star \citep{2017ARepSamus}. It has been included in CO line surveys of evolved stars, where molecular rotational lines are used to estimate circumstellar-envelope properties such as expansion velocity and mass-loss rate \citep{2010A&ADe}. Since G Her is a mass-losing semiregular AGB star, it is a useful target for testing whether the circumstellar envelope is spherical or affected by clumpy/asymmetric mass loss. The ProtoPol observations of G Her show modest continuum polarization, typically $\sim$ 0.3 - 0.8\%. The two epochs are broadly comparable, although the second epoch is slightly higher at some wavelengths. The behavior suggests a relatively weak but detectable asymmetric scattering environment \citep{1977A&ABrown}.

\paragraph{LQ Her} 
It is a late-type M giant with spectral type around M4.5 III and is a low-amplitude pulsating red giant \citep{2009MNRASTabur}. Long-term photometric monitoring shows that LQ Her exhibits multiple short pulsation periods, typical of small-amplitude semi-regular M giants rather than large-amplitude Mira variables \citep{2009MNRASTabur}. As such, LQ Her is likely associated with the upper-RGB or early-AGB population and is not expected to possess a dense dusty envelope. The ProtoPol observations of LQ Her show a clear epoch difference. The first epoch has stronger polarization and a noticeable increase toward longer wavelengths, reaching $>$ 1\%, while the second epoch remains lower and flatter. This is one of the clearer cases of wavelength-dependent and epoch-dependent continuum polarization, which would therefore most likely trace weak atmospheric asymmetry, molecular-layer scattering, surface inhomogeneity, interstellar polarization, or low-level circumstellar material.

\paragraph{$\omega$ Vir} 
It is a bright late-type M giant with spectral type around M4 III and is commonly classified as a low-amplitude semi-regular or irregular variable \citep{2017ARepSamus}. It is considered an evolved red giant, likely on or near the AGB, with reported light variations on short and long timescales, making it useful for studying weak pulsation and atmospheric extension in non-Mira red giants \citep{2003A&AKnapp, 2009MNRASTabur}. The ProtoPol observations of $\omega$ Vir show a strong epoch contrast. The first epoch has significantly higher polarization, often $\sim$ 1 - 2\%, while the second epoch remains much lower. The position angle also differs between epochs. This may suggest a substantial change in the scattering geometry or relative intrinsic/interstellar contribution. Since $\omega$ Vir is not a classical large-amplitude Mira or a strongly dust-enshrouded AGB star, the continuum polarization detected with ProtoPol would likely trace weak atmospheric asymmetry, molecular-layer scattering, surface inhomogeneity, interstellar polarization, or low-level circumstellar dust \citep{1977A&ABrown, maiti2026development}.

\paragraph{$\psi$ Vir} 
It is a bright evolved M giant with spectral type around M3 III and is generally classified as a low-amplitude irregular variable of Lb type \citep{2017ARepSamus}. Its small-amplitude variability and multiple reported pulsation timescales place it among the weakly pulsating red giants rather than the classical Mira variables \citep{2003A&AKnapp, 2009MNRASTabur}. For this reason, $\psi$ Vir is a useful comparison source for assessing whether low-amplitude red giants show intrinsic continuum polarization. The ProtoPol observations show low-to-moderate polarization, mostly below $\sim$ 0.8 \%, with only modest wavelength dependence. The two epochs are broadly similar, although the angle is scattered. This suggests weak intrinsic polarization or a relatively stable component.

\paragraph{R Gem} 
It is a classical long-period Mira variable and an evolved AGB star, with large-amplitude pulsations characteristic of Mira-type red giants \citep{2017ARepSamus, 2008MNRASWhitelock}. Mira variability is associated with strong radial pulsation, atmospheric shocks, extended molecular layers, and episodic dust formation, all of which can produce time-dependent asymmetries in the outer atmosphere and circumstellar envelope \citep{willson2000mass, 2018A&AHofner}. Therefore, R Gem is expected to be a stronger candidate for intrinsic continuum polarization than low-amplitude SRb or Lb giants. The ProtoPol observations of R Gem show moderate polarization with significant scatter and some localized deviations. The two epochs do not show a simple smooth trend, and $\theta$ is highly variable. Since R Gem is a Mira variable, such behavior may reflect phase-dependent atmospheric or circumstellar asymmetry, but the scatter suggests caution in interpretation \citep{1977A&ABrown, maiti2026development}.

\paragraph{RT Vir} 
It is an oxygen-rich semi-regular AGB star, commonly classified as an SRb variable, and is known to possess a circumstellar envelope traced by molecular and maser emission \citep{2017ARepSamus, 2010A&ADe}. VLBI observations of H$_2$O masers around RT Vir reveal an expanding, asymmetric circumstellar shell, indicating that its mass loss is not perfectly spherical \citep{2003ApJImai}. Long-term maser monitoring also shows temporal variability in the circumstellar gas, suggesting density inhomogeneities and evolving wind structure \citep{2020A&ABrand}. RT Vir is therefore a useful target for optical spectro-polarimetry, because any intrinsic continuum polarization detected with ProtoPol may trace clumpy or axisymmetric dust/molecular scattering associated with its non-spherical mass loss. The ProtoPol observations show relatively strong polarization at shorter wavelengths, especially in the first epoch, reaching up to $\sim$ 2 - 3\%, followed by a decline toward longer wavelengths. This strong wavelength dependence suggests scattering by circumstellar material and possible epoch-dependent changes in the envelope \citep{1977A&ABrown}.

\paragraph{RX Boo} 
It is a cool oxygen-rich semi-regular AGB star, commonly classified as an SRb variable \citep{2017ARepSamus}. It has been included in CO molecular-line surveys of evolved stars, where rotational transitions are used to estimate circumstellar-envelope properties such as expansion velocity and mass-loss rate \citep{2002A&AOlofsson, 2010A&ADe}. RX Boo is therefore a useful target for connecting optical continuum polarization with the structure of a known mass-losing AGB envelope. ProtoPol observations of RX Boo show moderate polarization, with the second epoch generally higher at shorter wavelengths and both epochs declining toward longer wavelengths. This behavior suggests a wavelength-dependent continuum-polarization component, likely associated with asymmetric circumstellar scattering.

\paragraph{ST UMa} 
It is a late-type semiregular red giant, commonly classified as an SRb variable with an M-type giant spectrum \citep{2017ARepSamus}. Semi-regular red giants such as ST UMa often show complex, low-amplitude, multiperiodic variability rather than a single stable Mira-like pulsation period \citep{1999A&AKiss, 2009MNRASTabur}. This makes ST UMa useful for studying weak atmospheric extension and possible low-level mass loss in non-Mira evolved giants. The ProtoPol observations show a clear epoch difference: the first epoch has stronger polarization, around $\sim$ 0.8 - 1.5\%, while the second epoch is lower. The polarization angle also differs between epochs. This points to variable intrinsic continuum polarization \citep{1977A&ABrown}.

\paragraph{SW Vir} 
It is a nearby oxygen-rich, non-Mira semiregular AGB star with spectral type M7 III and a distance of $143^{+19}_{-15}$ pc \citep{2019A&AOhnaka}. VLTI/AMBER observations spatially resolved the atmosphere of SW Vir in CO, H$_2$O, CN, and several atomic lines, showing that the star appears larger in molecular and atomic features than predicted by hydrostatic photospheric models \citep{2019A&AOhnaka}. The inferred H$2$O layer extends to about twice the stellar radius, making SW Vir a good target for testing whether extended molecular layers produce wavelength-dependent optical polarization \citep{2019A&AOhnaka}. The ProtoPol observations show one of the stronger and clearer epoch-dependent behaviors. The first epoch has higher polarization and increases toward longer wavelengths, while the second epoch remains lower. This suggests a strongly wavelength-dependent and time-variable scattering geometry, consistent with an asymmetric extended atmosphere or circumstellar envelope.

\paragraph{TU CVn } 
TU CVn is a late-type M giant, commonly listed as an M5 III semiregular variable with reported periods near 44.5 and 230 days \citep{2007MNRASGlass}. Its low-to-moderate variability places it among semiregular red giants rather than classical Mira variables, making it useful for probing weak atmospheric asymmetry and low-level mass loss \citep{1999A&AKiss, 2009MNRASTabur}. ProtoPol observations show continuum polarization for TU CVn \citep{maiti2026development}. A wavelength-dependent continuum polarization may therefore indicate scattering by asymmetric molecular or dusty material, while a weak or nearly flat component may include interstellar polarization \citep{maiti2026development}.

\paragraph{TV Gem} 
TV Gem is a bright late-type supergiant associated with the Gemini OB1 region at a distance of $\sim$ 1.4 - 1.5 kpc. It has a reported spectral type of M0 - M1.5 Iab and is typically classified as an SRc-type red supergiant variable rather than an AGB star. \citep{2017ARepSamus, 2019MNRASChatys}. Red supergiants commonly exhibit semi-regular or irregular variability caused by radial and non-radial pulsation, long secondary periods, and large convective surface structures \citep{2006MNRASKiss}. Therefore, any continuum polarization detected from TV Gem should be interpreted in the context of red-supergiant atmospheric asymmetry, large convective cells, and non-spherical mass loss, rather than Mira-like AGB pulsation. The ProtoPol observations of TV Gem show high continuum polarization, around $\sim$ 2 - 3\%, in both epochs, with comparatively small epoch-to-epoch changes. The polarization angle is also relatively stable. This stable, high polarization may include a significant interstellar component, although a stable asymmetric circumstellar/wind component cannot be ruled out \citep{1977A&ABrown, maiti2026development}. 

\paragraph{U Her} 
U Her is a classical oxygen-rich Mira variable on the AGB and is known to have a circumstellar maser shell \citep{2000A&AVan, 2017ARepSamus}. The OH 1667 MHz maser emission around U Her was observed using VLBA and has been used to estimate the astrometric position, proper motion, and yearly parallax of U Her \citep{2000A&AVan, 2003A&AVlemmings}. The first VLBA astrometric study reported a parallax of $5.3 \pm 2.1$ mas, corresponding to a distance of roughly $\sim$ 190 pc, although with a relatively large uncertainty \citep{2003A&AVlemmings}. The ProtoPol observations show strong polarization, with the first epoch much higher than the second epoch. The first epoch reaches $\sim$ 3\%, while the second epoch is closer to $\sim$ 1 - 1.5\%. This strong epoch dependence is consistent with variable intrinsic polarization, possibly related to Mira pulsation, shocks, or changing dust/molecular scattering regions.
\citep{willson2000mass, 2018A&AHofner, 1977A&ABrown, maiti2026development}. 

\paragraph{V636 Her} 
It is a late-type M giant, with a spectral type of M4.5 III, usually classified as a low-amplitude irregular or semi-regular variable \citep{2017ARepSamus}. In comparison with the Mira variables, V636 Her is expected to have less dramatic atmospheric extension and less mass-loss activity since it is a weakly pulsating red giant \citep{2003A&AKnapp}. V636 Her is therefore useful as a comparison source for assessing the lowest polarization levels in the red-giant sample. The ProtoPol observations show weak continuum polarization, generally below $\sim$ 1\%, with the first epoch slightly higher than the second epoch. The polarization angle is scattered, as expected for low $P$. V636 Her appears to have a weak intrinsic polarization signature. A weak or nearly wavelength-independent polarization would be consistent with a relatively symmetric atmosphere or interstellar contribution \citep{1977A&ABrown, maiti2026development}.

\paragraph{X Her } 
It is a nearby oxygen-rich semi-regular AGB star, located at a distance of roughly $\sim$ 130 -140 pc \citep{2022A&AAndriantsaralaza}. Molecular-line observations have shown that the CO profile of X Her has at least two kinematic components: a slowly expanding component associated with the circumstellar envelope and a faster component interpreted as an axisymmetric or bipolar outflow \citep{2008PASJNagakawa, 2011A&AJorissen}, indicating that its mass loss is not spherical. This makes X Her one of the most important targets in the sample for connecting optical continuum polarization with wind geometry. The ProtoPol observations of X Her show a very strong epoch contrast. The first epoch has high polarization, reaching $\sim$ 2 - 3 \%, while the second epoch is much lower, close to zero or below $\sim$ 0.5 \% over much of the wavelength range. This is one of the strongest cases for variable intrinsic continuum polarization in the sample. This may therefore indicate dust or molecular scattering associated with the asymmetric outflow, while epoch-to-epoch changes could indicate variable clumpy mass loss \citep{1977A&ABrown, maiti2026development}.
\par

\begin{figure*}
  \centering
  \includegraphics[angle=90, width=0.9\textwidth]{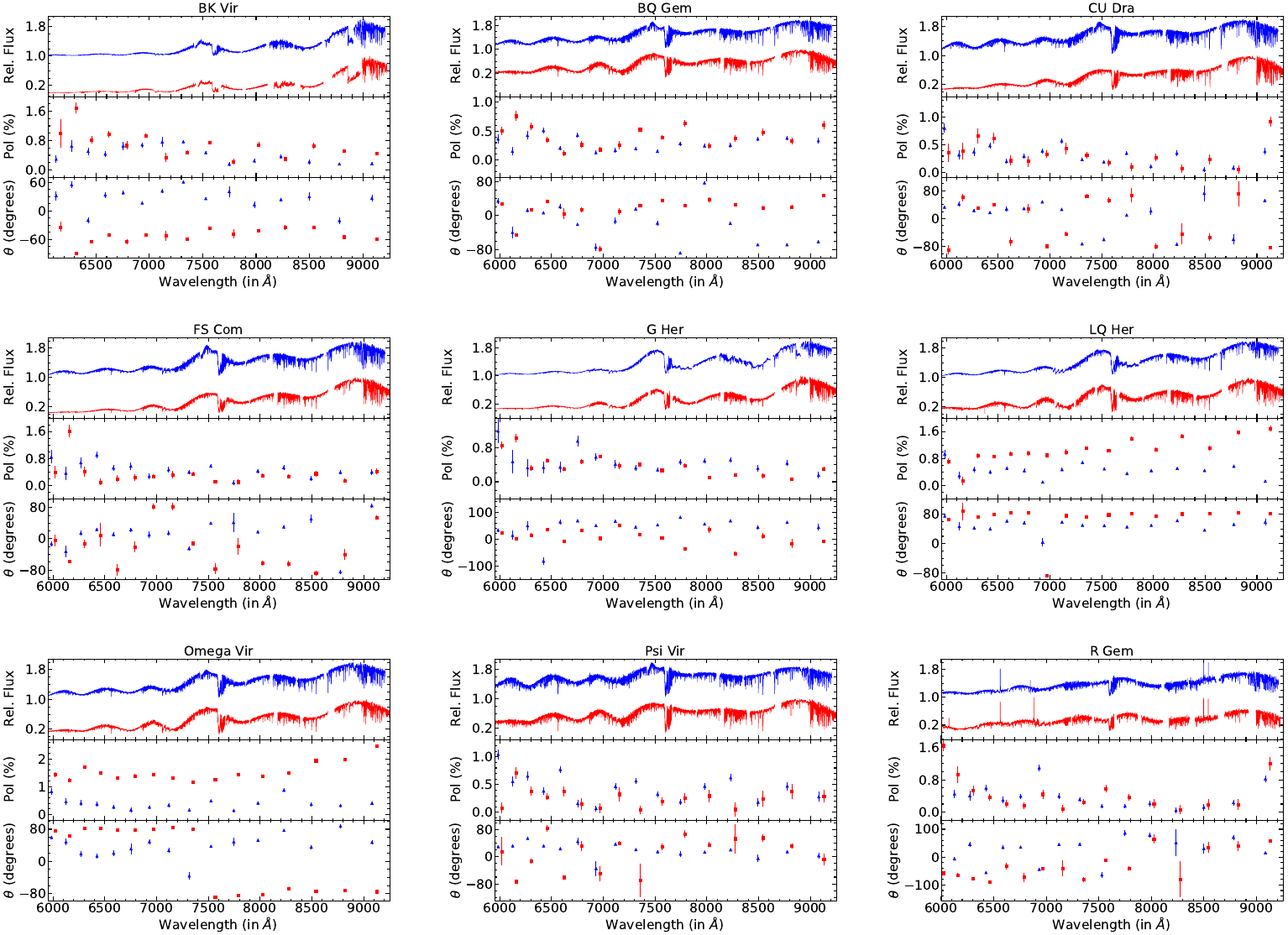}  
  \caption{Same as Figure~\ref{Fig-symbiotic_cont}, but for red giant stars BK Vir, BQ Gem, CU Dra, FS Com, G Her, LQ Her, Omega Vir, Psi Vir, and R Gem.}
  \label{Fig-AGB_1}
\end{figure*}

\begin{figure*}
  \centering
  \includegraphics[angle=90, width=0.9\textwidth]{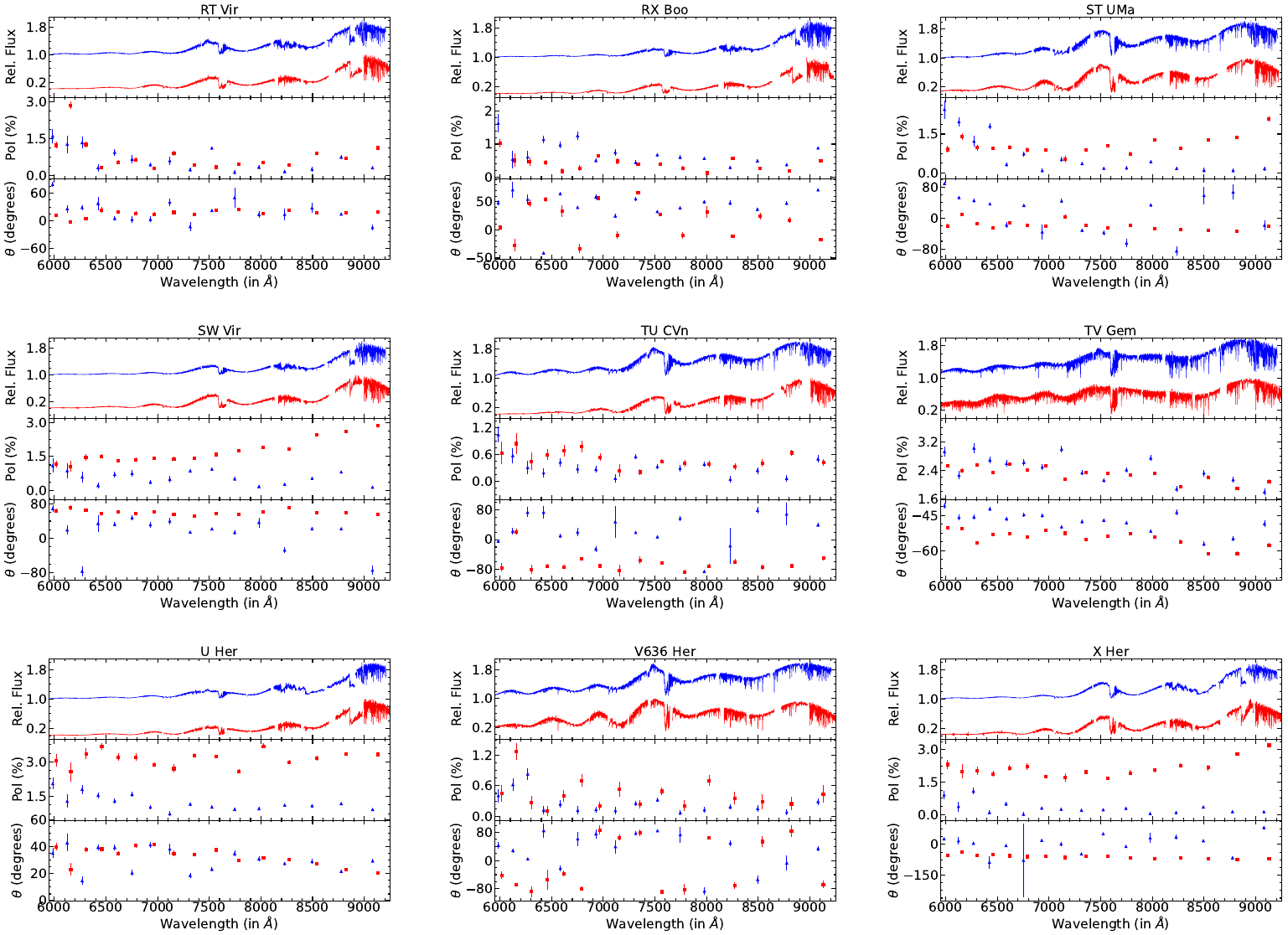}  
  \caption{Same as Figure~\ref{Fig-symbiotic_cont}, but for red giant stars RT Vir, RX Boo, ST UMa, SW Vir, TU CVn, TV Gem, U Her, V636 Her, and X Her.}
  \label{Fig-AGB_2}
\end{figure*}


\section{Summary and Conclusion} \label{sec:Summary}

We have presented a multi-epoch optical spectro-polarimetric study of evolved stellar systems observed with ProtoPol, a newly developed medium-resolution echelle spectro-polarimeter mounted on the PRL 2.5 m telescope at Mt. Abu Observatory. The observing campaign spans nearly 26 months, from March 2024 to May 2026, and includes 6 symbiotic stars and 18 red giant stars. The project evolved from the performance verification phase of the instrument, wherein a variety of objects were observed. Later,  as it was realized, the temporal variation in their spectra, if any, was sought to be explored to investigate continuum and line-dependent polarization variability. Identification of sources that show variability in their polarization spectra are of significance as they provide a suitable sample to study the unresolved circumstellar geometry, scattering regions, and temporal changes in the mass-loss environments of evolved stars.
\par
For the symbiotic-star sample, we examined both continuum polarization and polarization changes across key emission-line features, particularly H$\alpha$ and the Raman-scattered O~VI features at $\lambda\lambda$6830, 7088 {\AA}. Continuum polarization is detected in all observed systems, with several sources showing wavelength dependence or epoch-to-epoch variability. Since the measured continuum polarization may contain both intrinsic and interstellar components, the absolute values should be interpreted with caution. However, changes in the polarization level, wavelength dependence, or position angle between epochs indicate variable intrinsic scattering in at least some systems.
\par 
The H$\alpha$ spectro-polarimetric behavior varies across the sample. AG Peg and RW Hya show no clear polarization change across H$\alpha$, despite detectable continuum polarization. In contrast, UV Aur shows an enhancement in polarization across H$\alpha$, with a stronger line-polarization feature in the later epoch, suggesting an asymmetric and variable H$\alpha$ line-forming or scattering region. This polarization may arise from Raman scattering of Ly$\beta$ photons, Thomson scattering by free electrons, or a combination of both processes. T CrB shows strong variability in the H$\alpha$ intensity profile, including changes from single-peaked to double-peaked structures, but no significant polarization change across H$\alpha$ and no detectable Raman-scattered O~VI $\lambda\lambda$6830, 7088~\AA\ features. Its continuum polarization increases from $\sim$ 0.5\% to $\sim$ 1.1 \%, suggesting the possible development of an intrinsic polarization component during its recent active state. AG Dra and Z And show behavior consistent with complex Raman-scattering environments. In AG Dra, polarization across the Raman-scattered O VI features is variable, with a clear enhancement in the first epoch and little or no enhancement in the second, while H$\alpha$ shows no strong polarization change. In Z And, the continuum polarization is $\sim$1 - 2\%, with no significant increase in polarization degree across H$\alpha$ or the Raman features. However, a distinct rotation of the polarization angle is observed across the Raman features, indicating that the Raman-scattering region has a geometry different from the continuum-scattering region.
\par
For the red giant sample, the focus of this work is on continuum polarization. Measurable continuum polarization is detected in all 18 red giant targets, including semi-regular, irregular, Mira-type, and AGB stars, as well as the red supergiant TV Gem. Since an unresolved, spherically symmetric envelope should produce little or no net linear polarization, the detection of continuum polarization across the full sample indicates that weak asymmetry is common in the extended atmospheres or circumstellar environments of evolved cool stars. The continuum spectro-polarimetric behavior of the red giant sample shows substantial diversity across the two epochs. Measurable continuum polarization is detected in all targets, with amplitudes ranging from weak values of $\lesssim 0.5 \%$ in objects such as BQ Gem, CU Dra, and V636 Her, to strong values of $\sim2$ -3\% in RT Vir, SW Vir, TV Gem, U Her, and X Her. Several stars show clear epoch-to-epoch changes in the polarization degree, most notably LQ Her, $\omega$ Vir, ST UMa, SW Vir, U Her, and X Her. These variations indicate that at least part of the observed polarization is intrinsic and associated with changes in the circumstellar scattering geometry. In contrast, objects such as BQ Gem, CU Dra, G Her, $\psi$ Vir, and V636 Her show weaker polarization and less pronounced epoch-to-epoch variation, suggesting either relatively weak intrinsic asymmetry or a larger contribution from a stable interstellar component. The wavelength dependence of the polarization also varies across the sample. RT Vir, SW Vir, LQ Her, and X Her show clear wavelength-dependent polarization, consistent with scattering by circumstellar material in an asymmetric extended atmosphere or envelope. TV Gem shows consistently high polarization with comparatively little variation in either polarization degree or position angle between epochs, suggesting that a significant fraction of its observed polarization may arise from a stable interstellar or large-scale circumstellar component. In low-polarization sources, the polarization position angle is often scattered, as expected when the polarization amplitude is small and the angle becomes more uncertain. Overall, the two-epoch ProtoPol observations show that continuum polarization is common among evolved cool giants and that several objects exhibit temporal variability, supporting the presence of dynamic, non-spherical scattering environments.
\par
In this work, ProtoPol observations reveal the diagnostic potential of multi-epoch optical spectro-polarimetry for evolved stellar systems. For symbiotic stars, polarization in both lines and continuum constrains the geometry of neutral, ionized, and dusty scattering regions that are sculpted by binary interaction. In red giants, continuum polarization serves as a highly sensitive, indirect probe of subtle circumstellar asymmetries, even when the surrounding envelope is too compact or too faint to be spatially resolved. This study thus establishes ProtoPol and similar spectro-polarimeters as a powerful tool for tracking the time-variable scattering environments of evolved stars. Future observations with higher cadence, combined with information on orbital and pulsation phases, will be crucial for disentangling interstellar, instrumental, and intrinsic polarization contributions and for clarifying the physical mechanisms driving the observed variability.


\begin{acknowledgments}
The research work at the Physical Research Laboratory (PRL), Ahmedabad, is funded by the Department of Space (DOS), Govt. of India. AM and SSG gratefully acknowledge PRL for a Ph.D. research fellowship. RP gratefully acknowledges PRL for supporting her during her tenure at PRL through a post-doctoral fellowship from 9 January 2023 to 8 January 2025. PRL operates the Mt. Abu observatory with 1.2 m and 2.5 m telescopes at Mt. Abu. We acknowledge the use of data collected from both the PRL 1.2m and 2.5m telescopes at Mt. Abu Observatory with the ProtoPol instrument. This research has used the SIMBAD database, operated by CDS, Strasbourg, France.
\end{acknowledgments}

\begin{contribution}
AM was responsible for the observation, entire data reduction, and analysis with ProtoPol. He led the project with an initial research concept and manuscript preparation, particularly on the symbiotic star sample. RP was responsible for some of the first-epoch observations of red giant stars, manuscript preparation, editing, and discussion on the red giant sample of the manuscript. MKS is the PI of the ProtoPol instrument. He supervised the project and led the project with discussion, manuscript editing, etc. SSG was responsible for a few of the second epoch observations.
\end{contribution}

%
\facilities{PRL 1.2m and 2.5m telescopes at Mt. Abu Observatory, Gurushikhar, India}




\bibliography{Bibliography}{}
\bibliographystyle{aasjournalv7}



\end{document}